\newcommand{\CLASSINPUTbaselinestretch}{1.5}
\newcommand{\CLASSINPUTinnersidemargin}{1in}
\newcommand{\CLASSINPUTtoptextmargin}{1.1in}

\documentclass[onecolumn,12pt,draftcls]{IEEEtran}

\usepackage[dvips]{graphicx}
\usepackage[cmex10]{amsmath}
\usepackage{amssymb}
\usepackage{array}
\usepackage{enumerate}
\usepackage{psfrag}
\usepackage{stfloats}
\usepackage{subfigure}
\usepackage{dsfont}
\usepackage{algorithm}
\usepackage{algpseudocode}

\newcommand{\bbmat}{\begin{bmatrix} }
\newcommand{\ebmat}{\end{bmatrix} }

\graphicspath{{images/}}

\begin{document}
\title{Streaming Transmitter over Block Fading Channels with Delay Constraint}

\author{\IEEEauthorblockN{G. Cocco, D. G\"{u}nd\"{u}z and C. Ibars}\\
\IEEEauthorblockA{\small CTTC, Barcelona, Spain}\\
\small\{giuseppe.cocco, deniz.gunduz, christian.ibars\}@cttc.es
}

\date{}
\fontsize{12}{16}\selectfont
\maketitle
\vspace{-.8in}

\begin{abstract}
Data streaming transmission over a block fading channel is studied. It is assumed
that the transmitter receives a new message at each channel block at
a constant rate, which is fixed by an underlying application, and
tries to deliver the arriving messages by a common deadline. Various transmission schemes are proposed
and compared with an informed transmitter upper bound in terms of the average decoded rate. It is shown that in the single receiver case the adaptive joint encoding (aJE) scheme is asymptotically optimal, in that it achieves the ergodic capacity as the transmission deadline goes to infinity; and it closely follows the performance of the informed transmitter upper bound in the case of finite transmission deadline. On the other hand, in the presence of multiple receivers with different signal-to-noise ratios (SNR), memoryless transmission (MT), time sharing (TS) and superposition transmission (ST) schemes are shown to be more robust than the joint encoding (JE) scheme as they have gradual performance loss with decreasing SNR.
\end{abstract}
\vspace{-.4in}
\vspace{-.3in}
\section{Introduction}\label{sec:intro}
In a \emph{streaming transmitter} data becomes available over time rather than being available
at the beginning of transmission. Consider, for example, digital TV satellite broadcasting. The satellite receives video packets from a gateway on Earth at a fixed data rate and has to forward the received packets to the users within a certain deadline. Hence, the transmission of the first packet starts before the following packets arrive at the transmitter. We consider streaming transmission over a block fading channel with channel state information (CSI) available only at the receiver. This assumption results from practical constraints when the receiver belongs to a large population of terminals receiving a broadcast transmission, or when the transmission delay is significantly larger than the channel coherence time\footnote{Transmission rate can be adjusted to the channel state through adaptive coding and modulation (ACM) driven by a feedback channel. However, in real-time broadcast systems with large delays and many receivers, such as satellite systems, this is not practical. For instance, according to \cite{etsi_DVB_s2_user_guidelines} (Section 4.5.2.1) in real-time video transmission the ACM bit-rate control-loop may drive the source bit-rate (e.g., variable bit rate video encoder), but this may lead to a large delay (hundreds of milliseconds) in executing rate variation commands. In such cases the \mbox{total control loop delay is too large to allow real time compensation of fading.}} \cite{tse08_fading_BC_csi_rx}. The data that arrives at the transmitter over a channel block can be modeled as an independent
message whose rate is fixed by the quality of the gateway-satellite link and the video encoding scheme used for recording the event. We assume that the transmitter cannot modify the contents of the packets to change the data rate. This follows from the practical fact that the satellite transmitter is oblivious to the underlying video coding scheme adopted by the source, and considers the accumulated data over each channel block coherence time as a single data packet that can be either transmitted or dropped.
\begin{figure*}[]
\centering
\begin{small}
\psfrag{W1}{$W_1$}\psfrag{W2}{$W_2$}\psfrag{W3}{$W_3$}\psfrag{WM}{$W_M$}
\psfrag{t}{$t$}\psfrag{dead}{Deadline}
\psfrag{0}{$0$}\psfrag{n}{$n$}\psfrag{2n}{$2n$}\psfrag{M1}{$(M-1)n$}\psfrag{Mn}{$Mn$}
\psfrag{Ch1}{Ch. block 1}\psfrag{Ch2}{Ch. block 2}\psfrag{ChM}{Ch. block M}
\includegraphics[width=5in]{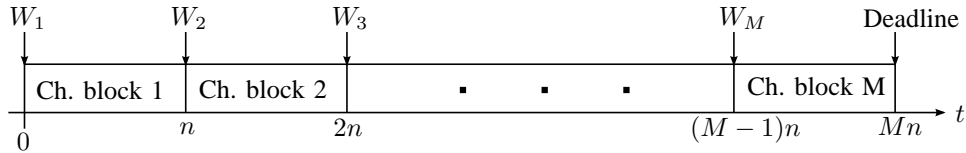} \caption{The transmitter receives message $W_i$ of rate $R$ at the beginning of channel block $i$. All the $M$ messages need to be transmitted to the receiver by the end of channel block $M$.}
\end{small}
\label{f:arrival}
\end{figure*}

We further impose a delay constraint on the transmission such that the receiver buffers the received messages for $M$ channel blocks before displaying the content, which is typical of multimedia streaming applications (see Fig. \ref{f:arrival}). As the messages arrive at the transmitter gradually over $M$ channel blocks, the last message sees only a single channel realization, while the first message can be transmitted over the whole span of $M$ channel blocks.
For a finite number $M$ of messages and $M$ channel blocks, it is not possible to average out the effect of fading in the absence of CSI at the transmitter, and there is always a non-zero outage probability \cite{Biglieri:IT:98}. Hence, the performance measure we study is the average decoded data rate by the user.

Communication over fading channels has been extensively studied \cite{Tse:book}. The capacity of a fading channel depends on the available information about the channel behavior \cite{Goldsmith:book}. When both the transmitter and the receiver have CSI, the capacity is achieved though waterfilling \cite{Goldsmith97_capac_fading}. This is called the ergodic capacity as the capacity is averaged over the fading distribution. In the case of a fast fading channel without CSI at the transmitter ergodic capacity is achieved with constant power transmission \cite{Tse:book}. However, when there is a delay requirement on the transmission as in our model, and the delay constraint is short compared to the channel coherence time, we have a slow fading channel. In a slow-fading channel, if only the receiver can track the channel realization, outage becomes unavoidable \cite{Tse:book}. An alternative performance measure in this case is the $\epsilon$-outage capacity \cite{Foschini:WPC:98}. In general it is hard to characterize the outage capacity exactly; hence, many works have focused on the high SNR \cite{Zheng:IT:03} or the low SNR \cite{Avestimehr:IT:07} asymptotic regimes. Another approach, which is also adopted in this work, is to study the average transmission rate as in \cite{Shamai:ISIT:97} and \cite{Shamai:IT:03}. Outages may occur even if the transmitter has access to CSI if it is required to sustain a constant transmission rate at all channel states. This can be due to the short-term power constraint, when the channel quality is so poor that the maximum power available is not sufficient to transmit the message reliably at the required rate \cite{Gunduz:TWC:07}; or, when the average power is not sufficient to sustain a constant rate at all channel conditions, which is called the delay-limited capacity \cite{Hanly:IT:98}.
Due to the constant rate of the arriving messages at all channel blocks our problem is similar to the delay-limited capacity concept. However, here we neither assume CSI at the transmitter nor require all arriving messages to be transmitted. Our work also differs from the average rate optimization in \cite{Shamai:ISIT:97} since the transmitter in \cite{Shamai:ISIT:97} can adapt the transmission rate based
on the channel characteristics and the delay constraint, whereas in our model the
message rate is fixed by the underlying application. The
degree-of-freedom the transmitter has in our setting is the multiple
channel blocks it can use for transmitting the messages while being
constrained by the causal arrival of the messages and the total
delay constraint of $M$ blocks.

Data streaming has received significant attention recently. Most of the work in this area focus on practical code construction \cite{Bogino:ISCAS:07}, \cite{Badr:Globecom:10}, \cite{Leong:ISIT:12}. More similar to our work, \cite{khisti_11_IT} studies the diversity-multiplexing tradeoff in a streaming transmission system
with a maximum delay constraint for each message. Unlike
in \cite{khisti_11_IT}, we assume that \emph{the whole} set of
messages has a common deadline; hence, in our setting the
degree-of-freedom available to the first message is higher than the
one available to the last.

In the present paper we extend our work in \cite{cocco11_real_time_BC} by presenting analytical results and introducing more effective transmission schemes.
We first study joint encoding (JE) which encodes all the
available messages into a single codeword at each channel block. We also study
time-sharing (TS) and superposition (ST) schemes. The main contributions of the present work can be summarized as follows:
\begin{enumerate}
  \item We introduce a channel model for streaming transmitter over block fading channels with a common decoding deadline to study real-time multimedia streaming in networks with large delays.
  \item We introduce an informed transmitter upper bound on the performance assuming the availability of perfect CSI at the transmitter.
  \item We show that a variant of the JE scheme, called the \emph{adaptive joint encoding (aJE)} scheme, performs very close to the informed transmitter upper bound for a finite number of messages, and approaches the ergodic capacity as the number of channel blocks goes to infinity.
  \item We show that the JE scheme has a phase transition behavior, which makes it unsuitable for networks with multiple receivers having different average SNRs. As an alternative, we propose the TS and ST schemes, whose performance degrade gradually with the decreasing average SNR.
\end{enumerate}
We support our analytical results with extensive numerical simulations. The rest of the paper is organized as follows. In Section
\ref{sec:sysmod} we describe the system model. In Section \ref{sec:Schemes} we describe the proposed transmission schemes in detail. In Section \ref{sec:upper_bound} we provide an informed transmitter upper bound on the average decoded rate, while Section \ref{sec:num_res} is devoted to the numerical results. Finally, Section \ref{sec:conclusions} contains the conclusions.
\vspace{-.3in}
\section{System Model}\label{sec:sysmod}
We consider streaming transmission over a block fading channel. The channel
is constant for a block of $n$ channel uses and changes in an
independent and identically distributed (i.i.d.) manner from one
block to the next. We assume that the transmitter accumulates the data that arrives at
a fixed rate during a channel block, and considers the accumulated data as a single message to be transmitted during the following channel blocks. We consider
streaming of $M$ messages over $M$ channel blocks, such that
message $W_t$ becomes available at the beginning of channel block
$t$, $t=1, \ldots, M$ (see Fig. \ref{f:arrival}). Each message $W_t$ has rate $R$ bits per
channel use (bpcu), i.e., $W_t$ is chosen randomly with uniform
distribution from the set $\mathcal{W}_t = \{1,\ldots,2^{nR}\}$, where $n$ is the number of channel uses per channel block. Following a typical assumption in the literature (see, e.g., \cite{Shamai:ISIT:97}), we assume that $n$, though still large (as to give rise to the notion of reliable communication
\cite{ozarow94_info_teo_mobile_cellular}), is much shorter than the dynamics of the slow fading
process. The channel in block $t$ is given by
\begin{eqnarray}\label{eqn:rx_signal}
\mathbf{y}[t] = h[t] \mathbf{x}[t] + \mathbf{z}[t],
\end{eqnarray}
where $h[t] \in \mathds{C}$ is the channel state, $\mathbf{x}[t] \in \mathds{C}^n$ is the
channel input, $\mathbf{z}[t] \in \mathds{C}^n$ is the i.i.d. unit-variance Gaussian noise, and $\mathbf{y}[t] \in \mathds{C}^n$ is the channel output.
The instantaneous channel
gains are known only at the receiver. We have a
short-term average power constraint of $P$, i.e.,
$E[\mathbf{x}[t]\mathbf{x}[t]^\dag] \leq nP$ for $t=1, \ldots, M$,
where $\mathbf{x}[t]^\dag$ represents the Hermitian transpose of
$\mathbf{x}[t]$ and $E[x]$ is the mean value of $x$. The short-term power constraint models the restriction on the maximum power radiated by the transmitter which is present in many practical systems \footnote{In cellular systems, for instance, the maximum power emitted by the transmitter is generally bounded in order to limit the interference to neighbor cells and keep it under a threshold value \cite{wireless_comm_book}. In satellite systems broadcasting multimedia traffic the onboard high power amplifier is generally driven to the limit of saturation in order to optimize the cost of the system by providing the maximum output power under given distortion constraints (\cite{maral_bousq_satcom_sys}, Section 9.2).}.

\begin{figure*}[]
\centering
\begin{tiny}
\psfrag{W1}{$W_1$}\psfrag{W2}{$W_2$}\psfrag{W3}{$W_3$}\psfrag{WM}{\hspace{-.1in}$W_M$}
\psfrag{Xt1}{$\mathbf{x}[1]$}\psfrag{Xt2}{$\mathbf{x}[2]$}\psfrag{Xt3}{$\mathbf{x}[3]$}\psfrag{XtM}{$\mathbf{x}[M]$}
\psfrag{h1}{$h[1]$}\psfrag{h2}{$h[2]$}\psfrag{h3}{$h[3]$}\psfrag{hM}{$h[M]$}
\psfrag{z1}{$\mathbf{z}[1]$}\psfrag{z2}{$\mathbf{z}[2]$}\psfrag{z3}{$\mathbf{z}[3]$}\psfrag{zM}{$\mathbf{z}[M]$}
\psfrag{y1}{$\mathbf{y}[1]$}\psfrag{y2}{$\mathbf{y}[2]$}\psfrag{y3}{$\mathbf{y}[3]$}\psfrag{yM}{$\mathbf{y}[M]$}
\psfrag{hW}{$(\hat{W}_1, \hat{W}_2, \ldots, \hat{W}_M)$}
\includegraphics[width=2.3in]{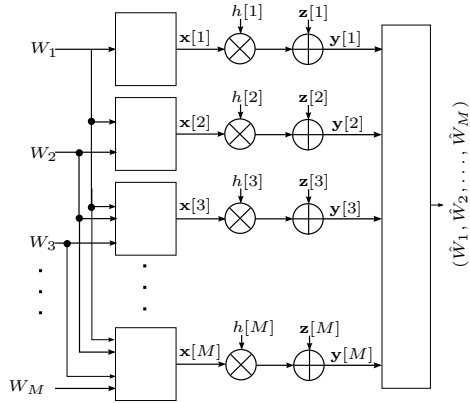} \caption{Equivalent channel
model for the sequential transmission of $M$ messages over $M$ channel blocks to a single receiver.}
\label{f:message}
\end{tiny}
\end{figure*}

The channel from the source to the receiver can be seen as a
multiple access channel (MAC) with a special message hierarchy
\cite{Prelov:PPI:84}, in which the encoder at each channel block
acts as a separate virtual transmitter (see Fig. \ref{f:message}), and the receiver tries to decode as many
of the messages as possible. Our performance measure is the average decoded rate. We denote the
instantaneous channel capacity over channel block $t$ by
$C_t \triangleq \log_2(1+ \phi[t]P)$, where $\phi[t]$ is a random
variable distributed according to a generic probability density function (pdf) $f_{\Phi}(\phi)$. Note that $C_t$ is also a random variable. We define $\overline{C}\triangleq E[\log_2(1+\phi
P)]$, where the expectation is taken over $f_{\Phi}(\phi)$. $\overline{C}$ is the ergodic capacity of this channel when there is no delay constraint on the transmission.
\vspace{-.3in}
\section{Transmission Schemes}\label{sec:Schemes}
The most straightforward transmission scheme is to send each message only within the channel block following its arrival. This is called memoryless transmission (MT). Due to the i.i.d. nature of the channel over blocks, successful decoding probability is constant over messages. Denoting this probability by $p \triangleq Pr\left\{C_t\geq R\right\}$, the probability that exactly $m$ messages are decoded is
\begin{eqnarray}\label{eqn:prob_1_decode_m}
\eta(m) \triangleq \binom{M}{m}p^m(1-p)^{M-m}.
\end{eqnarray}
Note that we have a closed-form expression for $\eta(m)$, and
it can be further approximated with a Gaussian distribution if we
let $M$ go to infinity, i.e.,
\begin{eqnarray}\label{eqn:norm_approx_m}
\eta(m) \simeq \frac{1}{\sqrt{2\pi M p(1-p)}}e^{-\frac{(m-M
p)^2}{2Mp(1-p)}}.
\end{eqnarray}
The average decoded rate of the MT scheme $\overline{R}_{MT}$ is found by evaluating $\sum_{m=1}^M m\eta(m)$. The MT scheme treats all messages equally. However, depending on the average channel conditions, it might be more beneficial to allocate more resources to some of the messages in order to increase the average decoded rate. In the following, we will consider three basic transmission schemes based on the type of resource allocation used. We will find the average decoded rate for these schemes and compare them with an upper bound that \mbox{will be introduced in Section \ref{sec:upper_bound}.}
\vspace{-.2in}
\subsection{Joint Encoding Transmission}\label{sec:joint}
In the \textit{joint encoding (JE)} scheme we generate a single multiple-index codebook for each channel block. For channel block $t$, we generate a $t$ dimensional codebook of size $s_1
\times \cdots \times s_t$, $s_i=2^{nR}, \ \forall
i\in\{1,\ldots,t\}$, with Gaussian distribution, and index the
codewords as $\mathbf{x}_t(W_1, \ldots, W_t)$ where $W_i \in \mathcal{W}=\{1,\ldots,2^{nR}\}$
for $i=1, \ldots, t$. The receiver uses joint typicality decoder and
tries to estimate as many messages as possible at the end of block
$M$. With high probability, it will be able to decode the first $m$
messages correctly if \cite{Prelov:PPI:84}:
\begin{align}\label{eqn:je_condition}
   (m-j+1)R & \leq \sum_{t=j}^m C_t, ~~~~ \forall~~j=1, 2, \ldots, m.
\end{align}

\begin{figure*}[]
\centering
\begin{tiny}
\psfrag{R}{$R$}\psfrag{2R}{$2R$}
\psfrag{C1}{$C_1$}\psfrag{C2}{$C_2$}\psfrag{0}{$0$}
\includegraphics[width=3.0in]{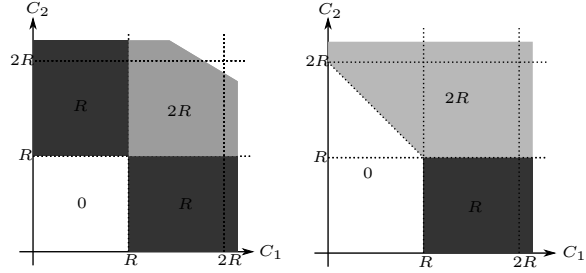}
\caption{Total decoded rate regions in the $(C_1, C_2)$ with $M=2$ messages for MT (on the left)
and JE (on the right) schemes.} \label{f:regions}
\end{tiny}
\end{figure*}

As a comparison, we illustrate the achievable rate regions for MT and JE schemes for $M=2$ in Fig.
\ref{f:regions}. In the case of MT, a total
rate of $2R$ can be decoded successfully if both capacities $C_1$ and $C_2$ are
above $R$. We achieve a total rate of $R$ if only one of the
capacities is above $R$. On the other hand, in the case of joint
encoding, we tradeoff a part of the region of rate $R$ for rate
$2R$; that is, we achieve a rate of $2R$ instead of rate $R$, while
rate $0$ is achieved rather than rate $R$ in the remaining region.

Using the conditions in (\ref{eqn:je_condition}) we define functions $f^m(R)$, for $m=0,1,\ldots,M$, as follows:
\begin{align}
f^m(R) &= \begin{cases}
1, & \text{ if } (m-j+1)R  \leq \sum_{t=j}^m C_t, j=1, \ldots, m, \\
0, & \text{ otherwise}. \nonumber
\end{cases}
\end{align}
Then the probability of decoding exactly $m$ messages can be written
as,
\begin{align}\label{eqn:decode_m}
   \eta(m) & = Pr\left\{ f^m(R)=1\text{ and } f^{m+1}(R)=0\right\}.
\end{align}
After some manipulation, it is possible to prove that exactly $m$
messages, $m=0,1,\ldots,M$, can be decoded if:
\begin{align}\label{eqn:je_cond_1}
    C_{m-i+1} + \cdots + C_m &\geq iR, ~~ i=1,\ldots, m, \\
    C_{m+1} + \cdots + C_{m+i} &< iR, ~~ i=1,\ldots, M-m. \label{eqn:je_cond_2}
\end{align}
\begin{figure*}[b!]
\hrulefill
\begin{small}
\begin{align}\label{eqn:decode_m_integral}
\eta(m)= & \int_{R}^{\infty}\int_{(2R-x_m)^+}^{\infty}
\cdots\int_{(mR-x_m-\cdots-x_2)^+}^{\infty}f_{C_1\cdots C_m}(x_1, \ldots, x_m)dx_1 \cdots dx_m \nonumber \\
& \times \int_{0}^{R}\int_{0}^{2R-x_{m+1}}
\cdots\int_{0}^{^{(M-m)R-x_{m+1}-\cdots-x_{M-1}}}f_{C_{m+1}\cdots
C_{M}}(x_{m+1}, \ldots, x_{M})dx_{m+1}\cdots dx_{M}
\end{align}
\end{small}
\end{figure*}
Then $\eta(m)$ can be calculated as in Eqn.
(\ref{eqn:decode_m_integral}) at the bottom of the page, where we
have defined $x^+=\max\{0,x\}$, and $f_{C_1\cdots C_m}(c_1, \ldots,
c_m)$ as the joint pdf of $C_1,\ldots ,C_m$, which is equal to
the product of the marginal pdf's due to independence. The
probability in Eqn. (\ref{eqn:decode_m_integral}) cannot be easily
evaluated for a generic $M$. However, we provide a much simpler way to
calculate the average
decoded rate $\overline{R}_{JE}$. The simplification of the average rate expression is valid not only for i.i.d. but also for \emph{conditionally i.i.d.} channels. Random variables $\{C_1,\cdots,C_M\}$ are said to be conditionally i.i.d. given a random variable $U$ if the joint distribution is of the form
\begin{eqnarray}\label{eqn:propos_condit_iid}
f_{C_1,\cdots,C_M,U}(c_1,\cdots,c_M,u)=f_{C_1|U}(c_1|u)\times\cdots \times f_{C_M|U}(c_M|u)f_U(u), ~~
\end{eqnarray}
where
\begin{eqnarray}\label{eqn:propos_condit_iid2}
f_{C_i|U}(c_i|u)=f_{C_l|U}(c_l|u),\ \forall i,l \in\{1,\ldots M\}.
\end{eqnarray}
Note that i.i.d. channels is a particular case of conditionally i.i.d. channels where $U$ is a constant.

\emph{Theorem 1:} The average decoded rate for the JE
scheme in the case of conditionally i.i.d. channel capacities is given by:
\begin{eqnarray}\label{eqn:theorem_1}
\overline{R}_{JE}=\frac{R}{M}\sum_{m=1}^{M}Pr\{C_1+\cdots +C_m\geq mR\}.
\end{eqnarray}

\emph{Proof:} See Appendix.

In general it is still difficult to find an exact
expression for $\overline{R}_{JE}$, but it is possible to show that
$\overline{R}_{JE}$ approaches $R$ for large $M$ if $\overline{C}> R$. To prove this, we rewrite Eqn. (\ref{eqn:theorem_1}) as:
\begin{align}\label{eqn:linearity_JE}
\overline{R}_{JE} = R - \frac{R}{M} \sum_{m=1}^{M}a_m,
\end{align}
where we have defined
\begin{align}\label{eqn:linearity_JE}
a_m\triangleq Pr\left\{\frac{C_1+\cdots+C_m}{m}<R\right\}.
\end{align}
It is sufficient to prove that, if $\overline{C}>R$, then $\lim_{M\rightarrow\infty}\sum_{m=1}^{M}a_m= c$, for some
$0<c<\infty$. We start by noting that $\lim_{m\rightarrow +\infty}a_m=0$,
since, by the law of large numbers, $\frac{C_1+\cdots +C_m}{m}$
converges to a Gaussian random variable with mean $\overline{C}$ and
variance $\frac{\sigma^2_c}{m}$ as $m$ goes to infinity, $\sigma^2_c$ being the variance of the channel capacity. To prove
the convergence of the series sum we show that
\begin{align}\label{eqn:ratio_serie}
\lim_{m\rightarrow +\infty}\frac{a_{m+1}}{a_m}=\lambda,
\end{align}
with $0<\lambda<1$.
We define
\begin{eqnarray}\label{eqn:define_lm}
l_m\triangleq\frac{\overline{C}-\frac{C_1+\cdots
+C_m}{m}}{\sigma_{c}/\sqrt{m}}, m=1, 2,\ldots, M,
\end{eqnarray}
where each $l_m$ is a random variable with zero mean
and unit variance. From the central limit theorem we can write:
\begin{eqnarray}\label{eqn:ratio_serie_clt}
\lim_{m\rightarrow +\infty}\frac{a_{m+1}}{a_m}&=&
\lim_{m\rightarrow+\infty}
\frac{Pr\left\{l_{m+1}>\frac{\overline{C}-R}{\sigma_{c}/\sqrt{m+1}}\right\}}
{Pr\left\{l_m>\frac{\overline{C}-R}{\sigma_{c}/\sqrt{m}}\right\}}\label{eqn:ratio_serie_clt_1}\\&=&
\lim_{m\rightarrow+\infty}\frac{Q\left(\frac{\overline{C}-R}{\sigma_{c}/\sqrt{m+1}}\right)}
{Q\left(\frac{\overline{C}-R}{\sigma_{c}/\sqrt{m}}\right)}\label{eqn:ratio_serie_clt_2}\\
&\leq& \lim_{m\rightarrow+\infty}\frac{
\frac{\sigma_c/\sqrt{m+1}}{(\overline{C}-R)\sqrt{2\pi}
}e^{-\frac{1}{2}\left(\frac{\overline{C}-R}{\sigma_c/\sqrt{m+1}}\right)^2}
  }{
  \frac{\frac{\overline{C}-R}{\sigma_c/\sqrt{m}}}{1+\left(\frac{\overline{C}-R}{\sigma_c/\sqrt{m}}\right)^2
}\frac{1}{\sqrt{2\pi}}e^{-\frac{1}{2}\left(\frac{\overline{C}-R}{\sigma_c/\sqrt{m}}\right)^2}
}\label{eqn:ratio_serie_clt_3}\\&=&\lim_{m\rightarrow+\infty}
\frac{\sigma_c^2+m(\overline{C}-R)^2}{\sqrt{m(m+1)}(\overline{C}-R)^2}
e^{-\frac{(\overline{C}-R)^2}{2}\left[\frac{m+1}{\sigma_c^2}-
\frac{m}{\sigma_c^2}\right]}\label{eqn:ratio_serie_clt_4}\\&=&e^{-\frac{(\overline{C}-R)^2}{2\sigma_c^2}} <1\label{eqn:ratio_serie_clt_5},
\end{eqnarray}
where inequality (\ref{eqn:ratio_serie_clt_3}) follows from the bounds on the Q-function:
\begin{eqnarray}\label{eqn:Q_bound}
\frac{x}{(1+x^2)\sqrt{2\pi} }e^{-\frac{x^2}{2}}< Q(x)<\frac{1}{x\sqrt{2\pi} }e^{-\frac{x^2}{2}}\ \mbox{ for}\ x>0. \blacksquare
\end{eqnarray}
In a
similar way, we prove that if $\overline{C}<R$, then the average
rate tends to zero asymptotically with $M$. To see this, we
consider the series in Eqn. (\ref{eqn:theorem_1}) defining $b_m=Pr\{C_1+\cdots +C_m\geq mR\}$. We want to prove that $\sum_{m=1}^{M}b_m$ converges to zero. We first notice that $\lim_{m\rightarrow +\infty}b_m=0$ by the
law of large numbers. Similarly to the above arguments, one can show
 that $\lim_{m\rightarrow +\infty}\frac{b_{m+1}}{b_m}=0$; and hence,
$\overline{R}_{JE}$ goes to zero as we increase the
number of messages and the channel blocks. Overall we see that the
average rate of the JE scheme shows a threshold behavior, i.e., we have:
\begin{eqnarray}\label{eqn:JEthreshold}
\lim_{M\rightarrow \infty}\overline{R}_{JE}=\begin{cases}
      R, & \mbox{if } R<\overline{C}   \\
      0, & \mbox{if } R>\overline{C}.
   \end{cases}
\end{eqnarray}

Eqn. (\ref{eqn:JEthreshold}) indicates a phase transition such that $\overline{R}_{JE}$ is zero even for large $M$ if $R > \bar{C}$ and the transmission rate cannot be modified. However, the transmitter may choose to transmit only a fraction $\alpha=\frac{M'}{M}<1$ of the messages, allocating the extra $M-M'$ channel blocks to the $M'$ messages, effectively controlling the transmission rate. In other words, the $M'$ messages are encoded and transmitted as described in the first part of this section in $M'$ channel blocks, while each of the remaining $M-M'$ blocks is divided into $M'$ equal parts, and the encoding process used for the first $M'$ blocks is repeated, using independent codewords, across the $M'$ parts of each block. For instance, let $M=3$ and $M'=2$. Then, $\mathbf{x}_1(W_1)$ and $\mathbf{x}_2(W_1,W_2)$ are transmitted in the first and second channel blocks, respectively. The third channel block is divided into $M'=2$ equal parts and the independent codewords $\mathbf{x}_{31}(W_1)$ and $\mathbf{x}_{32}(W_1,W_2)$ are transmitted in the first and in the second half of the block, respectively. We call this variant of the JE scheme \emph{adaptive JE (aJE)} scheme. The conditions for decoding exactly $m$ messages, $m=0,1,\ldots,M'$, in aJE can be obtained from those given in (\ref{eqn:je_cond_1}) and (\ref{eqn:je_cond_2}) by replacing $C_i$ with $C^*_i=C_i+\frac{1}{M'}\sum_{j=M'+1}^MC_j$, $i \in\{1,\ldots,M'\}$. Note that the random variables $C_i^*$, $i \in\{1,\ldots, M'\}$, are conditionally i.i.d., i.e., they are i.i.d. once the variable $U=\frac{1}{M'}\sum_{j=M'+1}^MC_j$ is fixed. This implies that Theorem 1 holds. In the following we prove that the average decoded rate of the aJE scheme $\overline{R}_{aJE}$ approaches $\alpha R$ for large $M$ if $\overline{C}>\alpha R$.
Similarly to the JE scheme, it is sufficient to prove that, if $\overline{C}> \alpha R$,
\begin{align}\label{eqn:linearity_aJE_2}
\lim_{M\rightarrow\infty}\sum_{m=1}^{M\alpha}a^*_m=c,
\end{align}
for some $0<c<\infty$, where $a^*_m\triangleq Pr\left\{\frac{C^*_1+\cdots +C^*_m}{m}<R\right\}$. We can rewrite $a^*_m$ as follows:
\begin{eqnarray}\label{eqn:aJE_rewrite_sum_term}
a^*_m&=&Pr\left\{\frac{C_1+\cdots+C_m +\frac{m}{M'}\sum_{j=M'+1}^MC_j}{m}<R\right\}\label{eqn:aJE_rewrite_sum_term_2}
\\&=&Pr\left\{\frac{C_1+\cdots+C_m}{m} + \frac{(1-\alpha)}{\alpha}\frac{1}{M(1-\alpha)}\sum_{j=M\alpha+1}^MC_j<R\right\}\label{eqn:aJE_rewrite_sum_term_4}
\\&=&Pr\left\{l_{m}>\frac{\overline{C}/\alpha-R}{\sigma_c\sqrt{\left(\frac{1}{m}+\frac{1-\alpha}{M\alpha^2}\right)}}\right\}\label{eqn:aJE_rewrite_sum_term_5},
\end{eqnarray}
where \begin{eqnarray}\label{eqn:define_lm_star}
l_m\triangleq\frac{\overline{C}/\alpha-\frac{C_1+\cdots
+C_m}{m}-\frac{(1-\alpha)}{\alpha}\frac{1}{M(1-\alpha)}\sum_{j=M\alpha+1}^MC_j}{\sigma_c\sqrt{\frac{1}{m}+\frac{1-\alpha}{M\alpha^2}}}
\end{eqnarray}
is a random variable with zero mean and unit variance.
Since $m<M$ and by the law of large numbers applied to Eqn. (\ref{eqn:aJE_rewrite_sum_term_5}) we find $\lim_{m\rightarrow +\infty}a^*_m=0$, since $l_m$
converges to a Gaussian random variable with zero mean and unit
variance as $m$ goes to infinity.
First we show that
\begin{eqnarray}\label{eqn:convergence_aJE_1}
\lim_{m\rightarrow +\infty}\left(\frac{a^*_m}{d_m}\right)=c',
\end{eqnarray}
for some $0<c'<+\infty$
where we have defined:
\begin{eqnarray}\label{eqn:define_d}
d_m\triangleq Pr\left\{l'_{m}>\frac{\overline{C}/\alpha-R}{\sigma_c\sqrt{\left(\frac{1}{m}+\frac{1-\alpha}{m\alpha^2}\right)}}\right\},
\end{eqnarray}
and
\begin{eqnarray}\label{eqn:define_l_prime}
l'_m\triangleq\frac{\overline{C}/\alpha-\frac{C_1+\cdots
+C_m}{m}-\frac{(1-\alpha)}{\alpha}\frac{1}{m(1-\alpha)}\sum_{j=M\alpha+1}^{M\alpha+m}C_j}{\sigma_c\sqrt{\frac{1}{m}+\frac{1-\alpha}{m\alpha^2}}}
\end{eqnarray}
such that $l'_m$ is a random variable with zero mean and unit variance. From Eqn. (\ref{eqn:convergence_aJE_1}) we find
\begin{eqnarray}\label{eqn:convergence_aJE_1_unroll}
\lim_{m\rightarrow +\infty}\left(\frac{a^*_m}{d_m}\right)&=&\lim_{m\rightarrow +\infty}\frac{Pr\left\{l_{m}>\frac{\overline{C}/\alpha-R}{\sigma_c\sqrt{\left(\frac{1}{m}+\frac{1-\alpha}{M\alpha^2}\right)}}\right\}}{Pr\left\{l'_{m}>\frac{\overline{C}/\alpha-R}{\sigma_c
\sqrt{\left(\frac{1}{m}+\frac{1-\alpha}{m\alpha^2}\right)}}\right\}}\label{eqn:convergence_aJE_1_unroll_1}\\
&=&\lim_{m\rightarrow +\infty}\frac{Q\left(\frac{\overline{C}/\alpha-R}{\sigma_c\sqrt{\left(\frac{1}{m}+\frac{1-\alpha}{M\alpha^2}\right)}}\right)}
{Q\left(\frac{\overline{C}/\alpha-R}{\sigma_c\sqrt{\left(\frac{1}{m}+\frac{1-\alpha}{m\alpha^2}\right)}}\right)}\label{eqn:convergence_aJE_1_unroll_2}\\
&\leq&\lim_{m\rightarrow +\infty}\frac{Q\left(\frac{\overline{C}/\alpha-R}{\sigma_c\sqrt{\left(\frac{1}{m}+\frac{1-\alpha}{m\alpha^2}\right)}}\right)}
{Q\left(\frac{\overline{C}/\alpha-R}{\sigma_c\sqrt{\left(\frac{1}{m}+\frac{1-\alpha}{m\alpha^2}\right)}}\right)}\label{eqn:convergence_aJE_1_unroll_3}\\&=&1\label{eqn:convergence_aJE_1_unroll_4},
\end{eqnarray}
where inequality (\ref{eqn:convergence_aJE_1_unroll_3}) follows from the fact that $m<M$ and from the fact that $Q(x)$ is monotonically decreasing in $x$. Then we show that
\begin{align}\label{eqn:convergence_d_aJE}
\lim_{M\rightarrow\infty}\sum_{m=1}^{M\alpha}d_m=c'',
\end{align}
for some $0<c''<+\infty$. To prove the convergence of the series sum we show that $\lim_{m\rightarrow +\infty}\frac{d_{m+1}}{d_m}=\lambda'$, for some $0<\lambda'<1$.
From the central limit theorem we can write:
\begin{eqnarray}\label{eqn:ratio_serie_clt_aJE}
\lim_{m\rightarrow +\infty}\frac{d_{m+1}}{d_m}&=&
\lim_{m\rightarrow+\infty}\frac{Pr\left\{l_{m+1}>\frac{\overline{C}/\alpha-R}{\sigma_c\sqrt{\left(\frac{1}{m+1}+\frac{1-\alpha}{(m+1)\alpha^2}\right)}}\right\}}
{Pr\left\{l_m>\frac{\overline{C}/\alpha-R}{\sigma_c\sqrt{\left(\frac{1}{m}+\frac{1-\alpha}{m\alpha^2}\right)}}\right\}}\label{eqn:ratio_serie_clt_aJE_1}\\&=&
\lim_{m\rightarrow+\infty}\frac{Q\left(\frac{\overline{C}/\alpha-R}{\sigma_c\sqrt{\left(\frac{1}{m+1}+\frac{1-\alpha}{(m+1)\alpha^2}\right)}}\right)}
{Q\left(\frac{\overline{C}/\alpha-R}{\sigma_c\sqrt{\left(\frac{1}{m}+\frac{1-\alpha}{m\alpha^2}\right)}}\right)}\label{eqn:ratio_serie_clt_aJE_2}\\
&\leq& \lim_{m\rightarrow+\infty}\frac{
\frac{\sigma_c\sqrt{\left(\frac{1}{m+1}+\frac{1-\alpha}{(m+1)\alpha^2}\right)}}{(\overline{C}/\alpha-R)\sqrt{2\pi}
}e^{-\frac{1}{2}\left(\frac{\overline{C}/\alpha-R}{\sigma_c\sqrt{\left(\frac{1}{m+1}+\frac{1-\alpha}{(m+1)\alpha^2}\right)}}\right)^2}
  }{
  \frac{\frac{\overline{C}/\alpha-R}{\sigma_c\sqrt{\left(\frac{1}{m}+\frac{1-\alpha}{m\alpha^2}\right)}}}{1+\left(\frac{\overline{C}/\alpha-R}{\sigma_c\sqrt{\left(\frac{1}{m}+\frac{1-
  \alpha}{m\alpha^2}\right)}}\right)^2
}\frac{1}{\sqrt{2\pi}}e^{-\frac{1}{2}\left(\frac{\overline{C}/\alpha-R}{\sigma_c\sqrt{\left(\frac{1}{m}+\frac{1-\alpha}{m\alpha^2}\right)}}\right)^2}
}\label{eqn:ratio_serie_clt_aJE_3}\\
&=&\lim_{m\rightarrow+\infty}\frac{\sqrt{\left(\frac{1}{m+1}+\frac{1-\alpha}{(m+1)\alpha^2}\right)}}{(\overline{C}/\alpha-R)^2}\frac{\left[\sigma_c^2\left(\frac{1}{m}+\frac{1-\alpha}{m\alpha^2}\right)+
(\overline{C}/\alpha-R)^2\right]}{\sqrt{\frac{1}{m}+\frac{1-\alpha}{m\alpha^2}}}e^{-\frac{(\overline{C}/\alpha-R)^2}{2\sigma^2_c}\left(\frac{\alpha^2}{\alpha^2-\alpha+1}\right)}\label{eqn:ratio_serie_clt_aJE_5}\\
&=&e^{-\frac{(\overline{C}/\alpha-R)^2}{2\sigma^2_c}\left(\frac{\alpha^2}{\alpha^2-\alpha+1}\right)} <1 \label{eqn:ratio_serie_clt_aJE_6},
\end{eqnarray}
where inequality (\ref{eqn:ratio_serie_clt_aJE_3}) follows from from the bounds on the Q-function given in Eqn. (\ref{eqn:Q_bound}).$\blacksquare$

From Eqn. (\ref{eqn:ratio_serie_clt_aJE_6}) it follows that $\lim_{M\rightarrow \infty}\overline{R}_{aJE}=R$ if $\alpha R<\overline{C}$. Similarly, it can be easily shown that $\lim_{M\rightarrow \infty}\overline{R}_{aJE}=0$ if $\alpha R>\overline{C}$. Thus by choosing $\alpha$ appropriately, we can have
\begin{eqnarray}\label{eqn:JEthreshold_adapt2}
\lim_{M\rightarrow \infty}\overline{R}_{aJE}= \min\{R, \bar{C}\}.
\end{eqnarray}
Eqn. (\ref{eqn:JEthreshold_adapt2}) suggests that the average transmission rate can be adapted at the message level while keeping a fixed rate at the physical layer. We will see in Section \ref{sec:upper_bound} that the maximum average decoded rate cannot be above this value; hence, as the number of messages and the channel blocks go to infinity, the aJE scheme achieves the optimal performance. We will show in Section \ref{sec:num_res} through numerical analysis that near optimality of the aJE scheme is valid even for finite $M$. However, we also note the threshold behavior of the performance of aJE; that is, when there are multiple users or inaccuracy in the channel statistics information at the transmitter, aJE performs very poorly for users whose average received SNR is below the target value. In the following we propose alternative transmission schemes with more gradual \mbox{performance change with the SNR}.
\vspace{-.5in}
\subsection{Time-Sharing Transmission (TS)}\label{sec:time_share}
One of the resources that the encoder can allocate among different
messages is the total number of channel uses within each channel
block. While the whole first channel block has to be dedicated to
message $W_1$ (the only available message), the second
channel block can be divided among the messages $W_1$ and $W_2$, and
so on so forth. Assume that the encoder divides the channel block
$t$ into $t$ portions $\alpha_{1t},\ldots , \alpha_{tt}$ such that
$\alpha_{it}\geq0$ and $\sum_{i=1}^t\alpha_{it}=1$. In channel block
$t$, $\alpha_{it}n$ channel uses are allocated to message $W_i$. A constant power $P$ is used throughout the block.
Then the total amount of received mutual information (MI) relative to
message $W_i$ is $I_{i}^{tot}\triangleq\sum_{t=i}^M\alpha_{it}C_t$. Letting $\alpha_{it}=1$ if $t=i$ and $\alpha_{it}=0$ otherwise, we obtain the MT scheme.

For simplicity, in the \emph{time sharing (TS)} scheme we assume equal
time allocation among all the available messages; that is, for
$i=1,\ldots,M$, we have $\alpha_{it}=\frac{1}{t}$ for
$t=i,i+1,\ldots,M$, and $\alpha_{it}=0$ for $t=1, \ldots, i$. The messages that arrive earlier are allocated more
resources; and hence, are more likely to be decoded. We have
$I_{i}^{tot}>I_{j}^{tot}$ for $1\leq i<j\leq M$. Hence, the
probability of decoding at least $m$ messages is:
\begin{eqnarray}\label{eqn:p_dec_m_super}
\varsigma(m) \triangleq Pr\{I_m^{tot}\geq R\}, \mbox{\ \ \ for\ } m=0, 1, \ldots, M,
\end{eqnarray}
where we define $I_{M+1}^{tot} = 0$ and
$I_0^{tot} = \infty$. Then the average decoded
rate is:
\begin{eqnarray}\label{eqn:rewrite_mutual_info_time div}
\overline{R}_{TS}&=&\frac{R}{M}\sum_{m=1}^{M} \varsigma(m) = \frac{R}{M}\sum_{m=1}^{M}Pr\left\{\frac{C_m}{m}+\frac{C_{m+1}}{m+1}+\cdots +
\frac{C_M}{M}\geq R\right\}.
\end{eqnarray}
\begin{figure}[]
\centering
\includegraphics[width=4in]{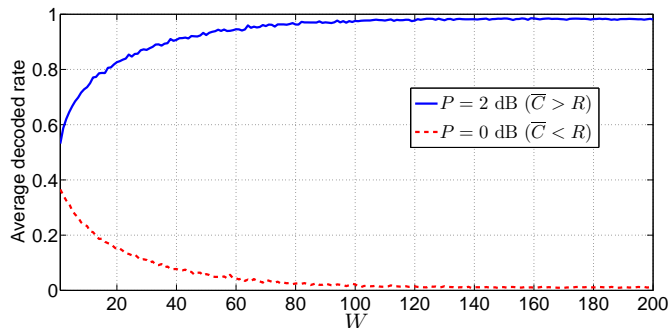}
\caption{Average decoded rate for the gTS scheme plotted against the
window size $W$ for $M=10^4$ messages and $R=1$ bpcu
for two different average SNR values.}
\label{fig:slide_TS_average_high_low}
\end{figure}
\vspace{-.5in}
\subsection{Generalized Time-Sharing Transmission (gTS)}\label{sec:g_time_share}
In \emph{generalized time-sharing} transmission each message is
encoded with equal time allocation over $W$ consecutive blocks as
long as the total deadline of $M$ channel blocks is not met. Messages from $W_1$ to $W_{M-W+1}$ are encoded over a
window of $W$ blocks, while messages $W_i$, for $i\in\{M-W+2, M-W+3, \ldots, M\}$
are encoded over $M-i+1$ blocks. In particular we focus on the
effect of variable $W$ on the average decoded rate $\overline{R}_{gTS}$. In case $W\ll M$
and $W\gg1$, most of the messages are transmitted over $W$ slots
together with $W-1$ other messages. In this case the MI
accumulated for a generic message $W_i$ is:
\begin{eqnarray}\label{eqn:tot_mutual_info_time div_windowed}
I_{i}^{tot}=\frac{1}{W}\sum_{t=i}^{i+W-1}C_t.
\end{eqnarray}
By the law of large numbers, (\ref{eqn:tot_mutual_info_time
div_windowed}) converges in probability to the average channel
capacity $\overline{C}$ as $W\rightarrow \infty$.
 Thus, we expect that, when the transmission
rate $R$ is above $\overline{C}$, the gTS scheme shows poor
performance for large $W$ (and hence, large $M$), while almost all
messages are received successfully if $R<\overline{C}$. We confirm
this by analyzing the effect of $W$ on $\overline{R}$ numerically in Fig. \ref{fig:slide_TS_average_high_low} for $M=10^4$ and
$R=1$ bpcu. For $P=0$ $\mathrm{dB}$ the average channel capacity $\overline{C}$ is
lower than $R$, which leads to a decreasing
$\overline{R}_{gTS}$ with increasing window size $W$. On the other hand,
for $P=2$ $\mathrm{dB}$ $\overline{C}$ is higher than
$R=1$ bpcu, and accordingly $\overline{R}_{gTS}$ approaches $1$ as
$W$ increases.

\begin{figure}[]
     \begin{center}
        \subfigure[$R=1$ bpcu ($R<\overline{C}$).]{%
            \label{fig:slide_comparison}
            \includegraphics[width=0.5\textwidth]{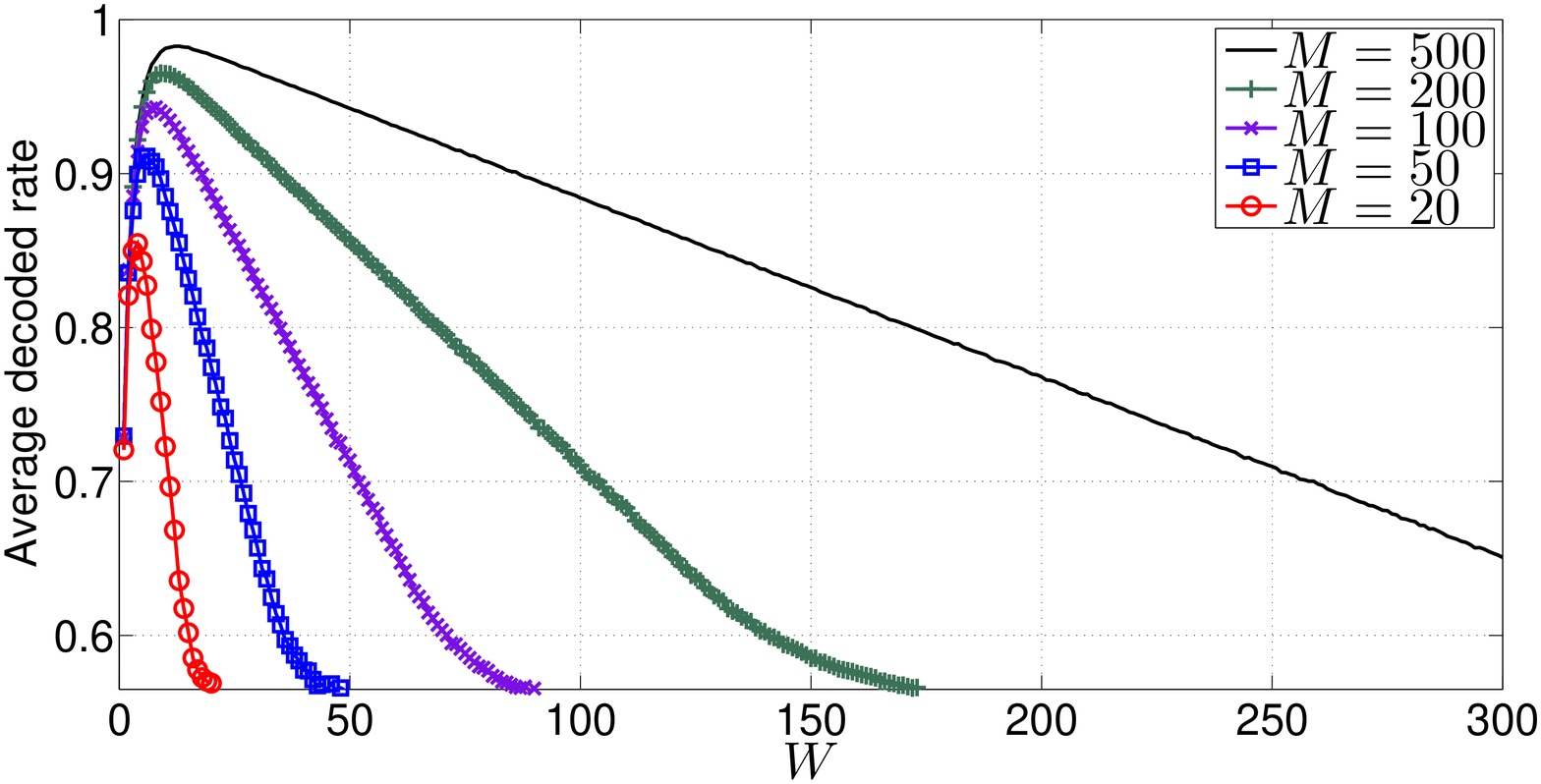}
        }%
        \subfigure[$R=1$ bpcu ($R>\overline{C}$).]{%
           \label{fig:slide_comparison_low_snr}
           \includegraphics[width=0.5\textwidth]{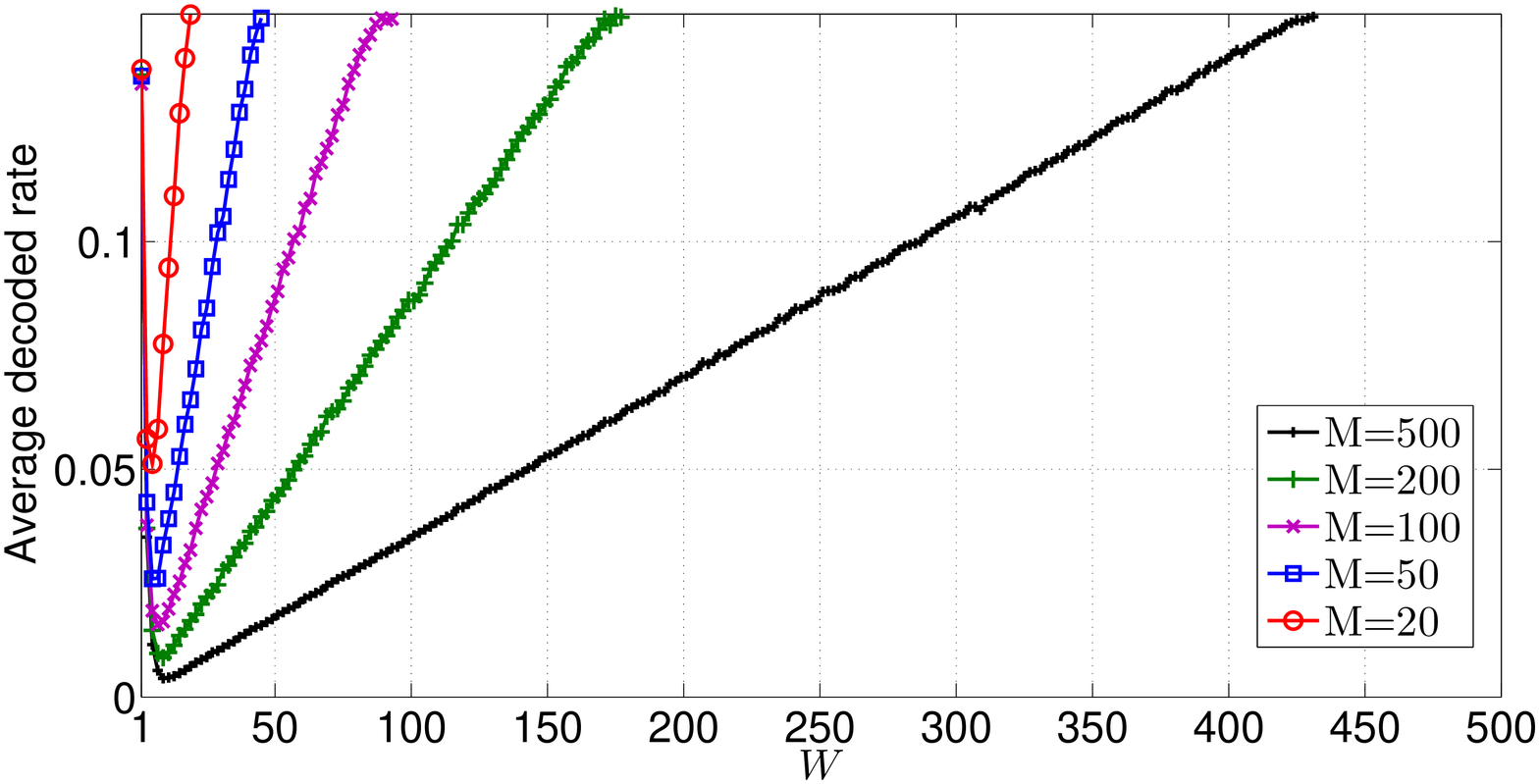}
        }\\ 
    \end{center}
    \caption{%
        Average decoded rate for the gTS scheme plotted against the window size $W$ for different values of $M$, $P=5$ $\mathrm{dB}$.
     }%
   \label{fig:subfigures}
\end{figure}
The same reasoning cannot be applied if the window size is of the
order of the number of messages, as the number of initial messages
which share the channel with less than $W-1$ other messages and the
number of final messages which share the channel with more than
$W-1$ messages are no longer negligible with respect to $M$. In Fig. \ref{fig:slide_comparison}, we plot $\overline{R}_{gTS}$ vs $W$ for relatively small numbers of
messages and $\overline{C}\geq R$. As seen in the figure,
for a given value of $M$ an optimal value of $W$ can be chosen to
maximize $\overline{R}_{gTS}$. The optimal value of $W$ increases with $M$ when $R < \bar{C}$. We plot $\overline{R}_{gTS}$ for
$\overline{C}<R$ in Fig.
\ref{fig:slide_comparison_low_snr}. From the figure we see that $\overline{R}_{gTS}$ decreases monotonically with $W$ up to a
minimum, after which it increases almost linearly. The initial
decrease in the decoded rate is due to the averaging effect
described above, while the following increase is due to the fact
that messages which are transmitted earlier get an increasing amount
of resources as $W$ increases, and so the probability to be decoded
increases. As a matter of fact, for each finite $i$, the average
MI accumulated for message $i$ grows indefinitely
with $W$, i.e.:
$$\lim_{W\rightarrow \infty} E\left\{\sum_{t=i}^{i+W-1}\frac{C_t}{t}\right\}\notag\\
=\lim_{W\rightarrow \infty}
\overline{C}\sum_{t=i}^{i+W-1}\frac{1}{t}=+\infty.$$ Thus, for a
fixed $i$, letting $W$ go to infinity leads to an infinite average
MI, which translates into a higher $\overline{R}_{gTS}$. Note that this is valid only for relatively small $i$ and
large $W$, i.e., only messages transmitted earlier get advantage from
increasing $W$, while the rest of the messages are penalized. For
instance, if $M>W$, while message $W_1$ is allocated a total of $n\sum_{t=1}^{W}\frac{1}{t}$ channel uses over $W$ channel blocks, message $W_M$ only receives a fraction
$\frac{1}{W}$ of a channel block. If $W$ is
small compared to $M$, as in the plot of Fig.
$\ref{fig:slide_TS_average_high_low}$ for $P=0$ $\mathrm{dB}$, the fraction of messages which get advantage from the
increasing $W$ remains small compared to $M$; and hence, $\overline{R}_{gTS}$ does not increase with $W$ for the considered range.

Note that the TS scheme in Section \ref{sec:time_share} is a special case of the gTS scheme obtained by letting $W=M$. On the
other extreme, by letting $W=1$, we obtain the MT scheme.

Although the idea of encoding a message over a fraction of the available consecutive slots (e.g., $W<M$ for message $W_1$ in gTS)
can be applied to all the schemes considered in this paper, the
analysis becomes quite cumbersome. Hence, we restrict our analysis to the TS scheme as explained above.
\vspace{-.2in}
\subsection{Superposition Transmission (ST)}\label{sec:sup_coding}
Next we consider \emph{superposition transmission (ST)}, in which
the transmitter transmits in channel block $t$, $t\in\{1,\ldots,M\}$, the superposition of $t$
codewords, chosen from $t$ independent Gaussian codebooks of size
$2^{nR}$, corresponding to the available messages $\{W_1, \ldots,
W_t\}$. The codewords are scaled such that the average total
transmit power in each block is $P$. In the first block, only
information about message $W_1$ is transmitted with average power
$P_{11}=P$; in the second block we divide the total power $P$ among
the two messages, allocating $P_{12}$ and $P_{22}$ for $W_1$ and
$W_2$, respectively. In general, over channel block $t$ we allocate
an average power $P_{it}$ for $W_i$, while $\sum_{i=1}^t P_{it}=P$.

Let $\mathcal{S}$ be any subset of the set of messages $\mathcal{M}
= \{1, \ldots, M\}$. We define $C(\mathcal{S})$ as follows:
\begin{eqnarray}\label{eqn:optimal_superposition_0}
    C(\mathcal{S}) \triangleq \sum_{t=1}^M \log_2 \left(1 + \frac{\phi[t]\sum_{s \in \mathcal{S}} P_{st}}{1 + \phi[t]\sum_{s \in \mathcal{M}
    \backslash \mathcal{S}} P_{st}} \right).
\end{eqnarray}
This provides an upper bound on the total rate of messages in set
$\mathcal{S}$ that can be decoded jointly at the user considering
the codewords corresponding to the remaining messages as noise.
The receiver first checks if any of the messages can be decoded
alone by considering the other transmissions as noise. If a message
can be decoded, the corresponding signal is subtracted and the
process is repeated over the remaining signal. If no message can be
decoded alone, then the receiver considers joint decoding of message
pairs, followed by triplets, and so on so forth. This optimal
decoding algorithm for superposition transmission is outlined in
Algorithm \ref{a:total_rate} below. The user calls the algorithm
with $Rate=0$ and $\mathcal{M}=\{1, \ldots, M\}$ initially.
\begin{algorithm}
\begin{small}
\caption{Total\_Decoded\_Rate ($Rate$, $\mathcal{M}$, $\mathbf{P}$)}
\label{a:total_rate}
{\begin{algorithmic} \State boolean
$Decoded=0$ \For{$i=1$ to $|\mathcal{M}|$}
                \If{$iR \leq \max_{\mathcal{S}: \mathcal{S} \subseteq \mathcal{M}, |\mathcal{S}|=i}
                C(\mathcal{S})$}
                \State $Decoded=1$
                 \State $Rate = Rate + \frac{iR}{M}$
                 \State $\mathcal{M} = \mathcal{M} \backslash \mathcal{S}$
                 \State \textbf{quit for}
                 \EndIf
                 \EndFor

             \If {($\mathcal{M} \neq \emptyset$) AND ($Decoded$)}
            \State\verb"Total_Decoded_Rate" ($Rate$, $\mathcal{M}$, $\mathbf{P}$)
            \Else
            \Return $Rate$
            \EndIf
            \end{algorithmic}}
\end{small}
\end{algorithm}

While Algorithm \ref{a:total_rate} gives us the maximum total rate,
it is challenging in general to find a closed form expression for
the average total rate, and optimize the power allocation. Hence, we focus here on the special case of equal
power allocation, where we divide the total average power
$P$ among all the available messages at each channel block.
The performance of the ST scheme will be studied in Section \ref{sec:num_res} numerically and compared with the other transmission schemes and an upper bound which will be introduced next.
\vspace{-.2in}
\section{Upper Bound}\label{sec:upper_bound}
We provide an upper bound on the performance by
assuming that the transmitter is informed about the exact channel
realizations at the beginning of the transmission. This allows the transmitter to
optimally allocate the resources among messages to maximize $\overline{R}$. Assume that $C_1,\ldots, C_M$ are known by the transmitter and the maximum
number of messages that can be decoded is $m\leq M$. We can always have the first $m$ messages to be the
successfully decoded ones by reordering. When the channel state is
known at the transmitter, the first $m$ messages can be decoded
successfully if and only if \cite{Prelov:PPI:84},
\begin{align*}
    iR & \leq C_{m-i+1} + C_{m-i+2} + \cdots + C_M, \mbox{\ for\ } i=1, \ldots, m.
\end{align*}
We can equivalently write these conditions
as
\begin{align}\label{cond_ub}
    R \leq \min_{i\in\{1, \ldots, m\}}\left[\frac{1}{m-i+1} \sum_{j=i}^M C_j\right].
\end{align}
Then, for each channel realization $\{h[1],\ldots,h[M]\}$, the upper bound on the average
decoded rate is given by $\frac{m^*}{M}R$, where $m^*$ is the greatest $m$
value that satisfies (\ref{cond_ub}). This is an upper bound for each specific channel realization obtained by optimally allocating the resources. An upper bound on $\overline{R}$ can be obtained by averaging this over the distribution of the channel realizations.

Another upper bound on $\overline{R}$ can be found from the ergodic capacity assuming all messages are available at the encoder at the beginning and letting $M$ go to infinity. \mbox{Thus, $\overline{R}$ can be bounded as:}
\begin{eqnarray}\label{eqn_looser_ub}
\overline{R}\leq \min\left\{R,\overline{C}\right\}.
\end{eqnarray}
The bound $\overline{R}\leq R$ follows naturally from the data arrival rate. Comparing (\ref{eqn_looser_ub}) and (\ref{eqn:JEthreshold_adapt2}) we see that the aJE scheme achieves the optimal average decoded rate in the limit of infinite $M$.
\vspace{-.2in}
\section{Numerical Results}\label{sec:num_res}
\begin{figure}[]
\centering
\includegraphics[width=4in]{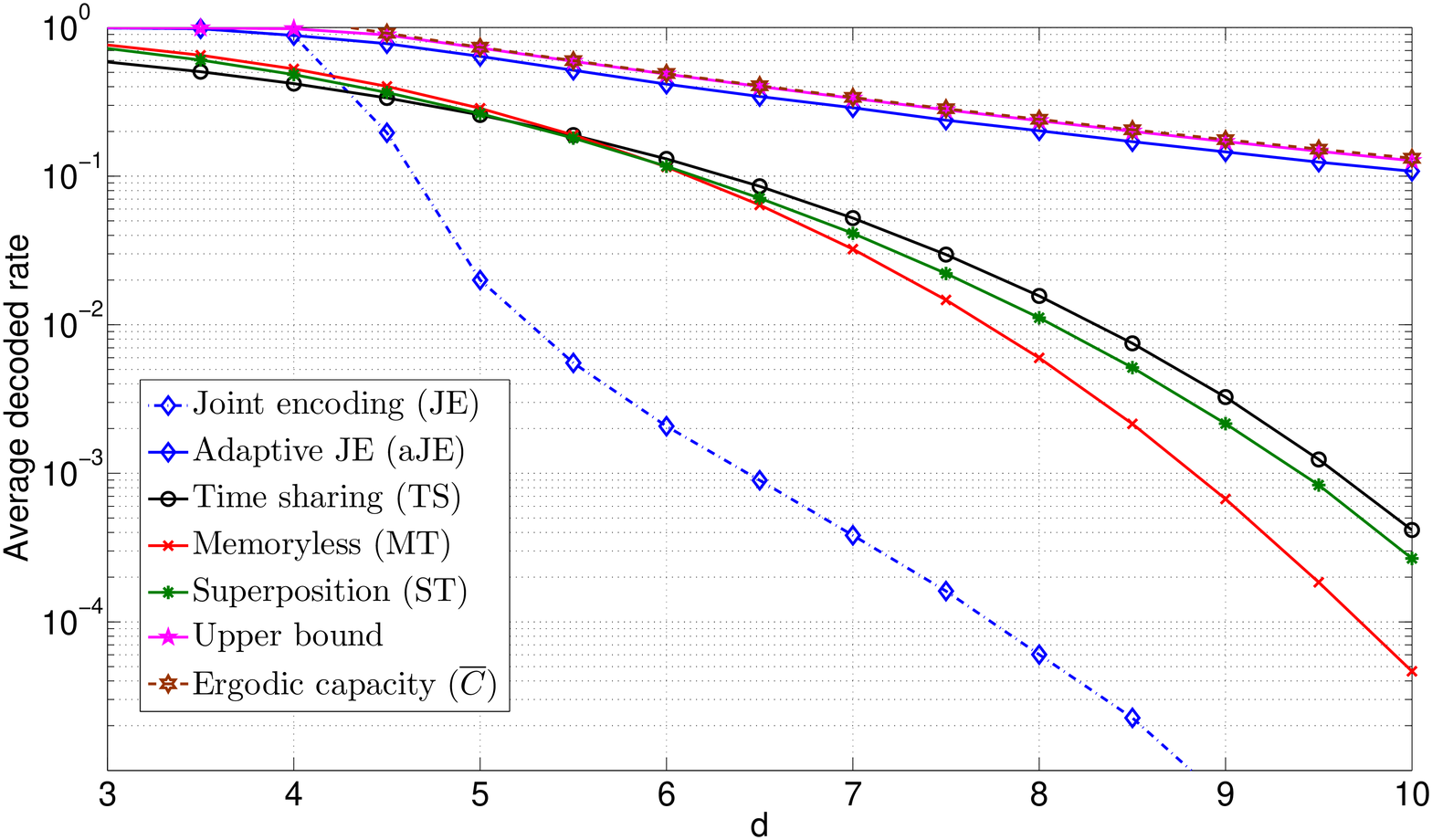}
\caption{Average decoded rate $\overline{R}$ vs
distance from the transmitter for $R=1$ bpcu, $M=100$, $P=20$ $\mathrm{dB}$ and $\alpha=3.$}
\label{fig:distance_comparison}
\end{figure}

In this section we provide numerical results comparing the
proposed transmission schemes. For the simulations we assume that the channel is Rayleigh fading, i.e., the channel state $\phi(t)$ is exponentially distributed with parameter $1$, i.e., $f_{\Phi}(\phi)=e^{-\phi}$ for $\phi >0$, and zero otherwise.
\begin{figure}[h!]
     \begin{center}
        \subfigure[$P=1.44$ $\mathrm{dB}$ ($\overline{C}> R$).]{%
            \label{fig:cumulative_j_decode_m_all}
            \includegraphics[width=0.5\textwidth]{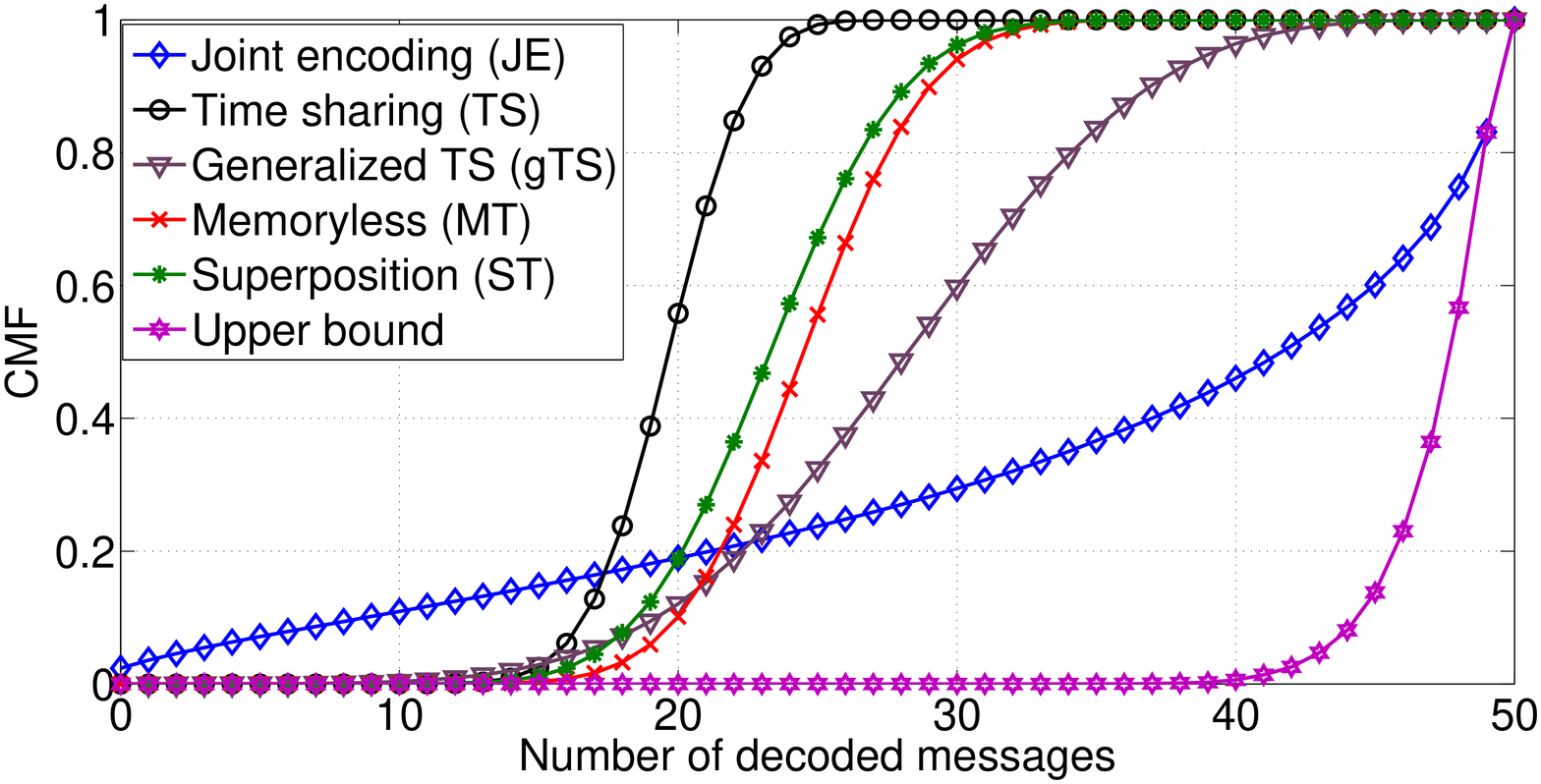}
        }%
        \subfigure[$P=0$ $\mathrm{dB}$ ($R>\overline{C}$).]{%
           \label{fig:cmf_low_snr}
           \includegraphics[width=0.5\textwidth]{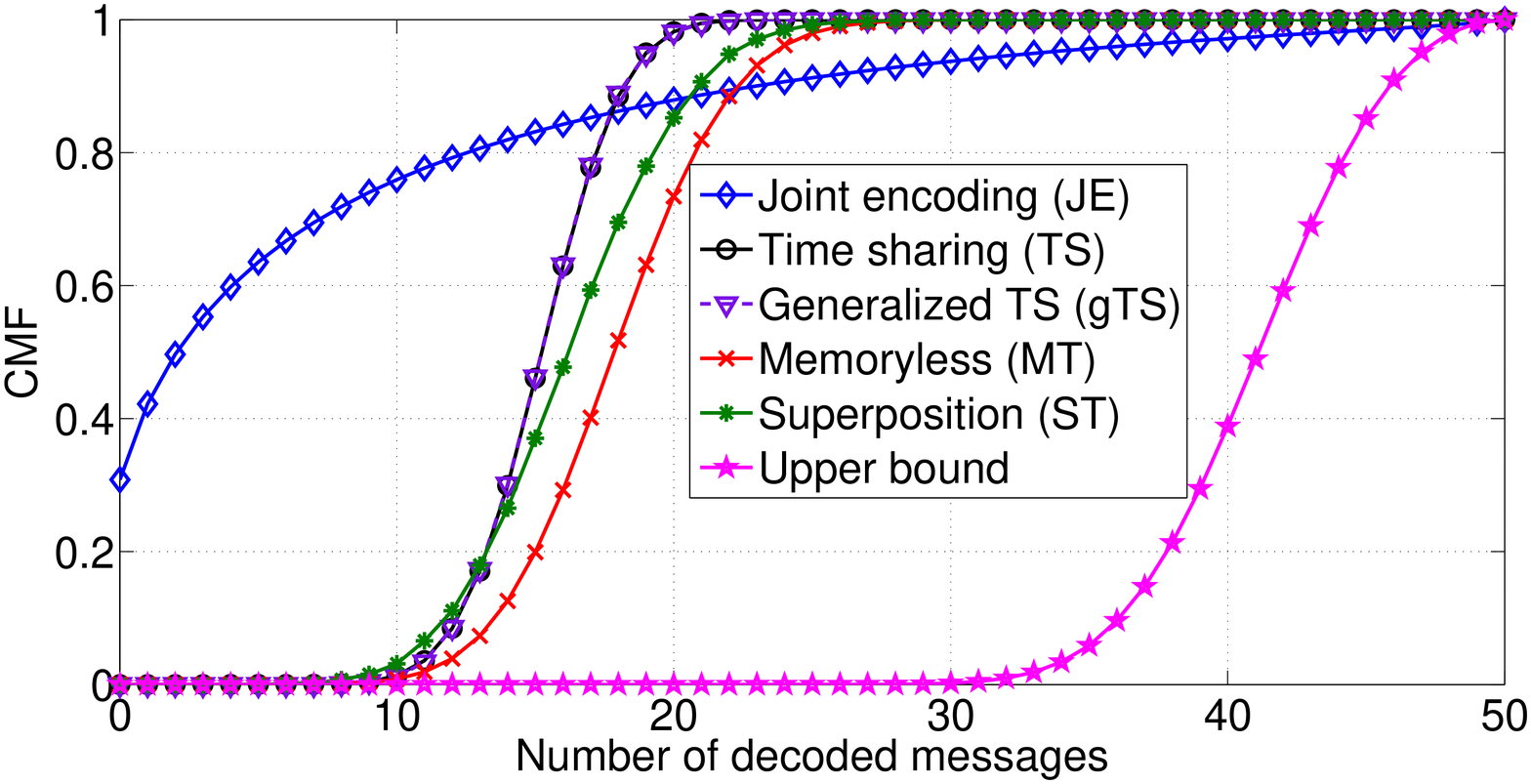}
        }\\
    \end{center}
    \caption{%
        The cumulative mass function (cmf) of the number of decoded messages for $R=1$ bpcu and $M=50$.
     }%
   \label{fig:subfigures}
\end{figure}
In Fig. \ref{fig:cumulative_j_decode_m_all} the cumulative mass function (cmf) of
the number of decoded messages is shown for the different transmission techniques
for $R=1$, $M=50$ and $P=1.44$ $\mathrm{dB}$, which corresponds to an
outage probability of $p=0.5$ for the MT scheme and an average
channel capacity $\overline{C}\simeq1.07 >R$. We see that MT
outperforms ST and TS schemes, as its cmf lays below the other two.
On the other hand, the comparison with the JE scheme depends on the performance metric we choose. For
instance, JE has the lowest probability to decode more than $m$
messages, for $m\leq15$, while it has the highest
probability for $m\geq 22$.
In Fig. \ref{fig:cmf_low_snr} the cmf's for the case of $P=0$ $\mathrm{dB}$
are shown. In this case the average capacity is $\overline{C}\simeq0.86$. Comparing Fig. \ref{fig:cmf_low_snr} and Fig. \ref{fig:cumulative_j_decode_m_all}, we see how the cmf of the JE scheme
has different behaviors depending on whether $\overline{C}$ is
above or below $R$. We see from Fig. \ref{fig:cmf_low_snr}
that for the JE scheme there is a probability of about $0.3$ not to
decode any message, while in all the other schemes such probability
is zero. However, the JE scheme also has the highest probability to
decode more than $30$ messages. Furthermore, we note that the cmf of the
gTS scheme converges to the cmf of TS scheme at low SNR. This is
because, as shown in Section \ref{sec:g_time_share}, when $\overline{C}<R$, the optimal window
size $W$ is equal to $M$, which is
nothing but the TS scheme. In the following, we focus on
the average decoded rate as our performance metric.
\begin{figure}[t!]
     \begin{center}
        \subfigure[$P=-3$ $\mathrm{dB}$ ($R>\overline{C}$).]{%
            \label{fig:rate_vs_M_0_db}
            \includegraphics[width=0.5\textwidth]{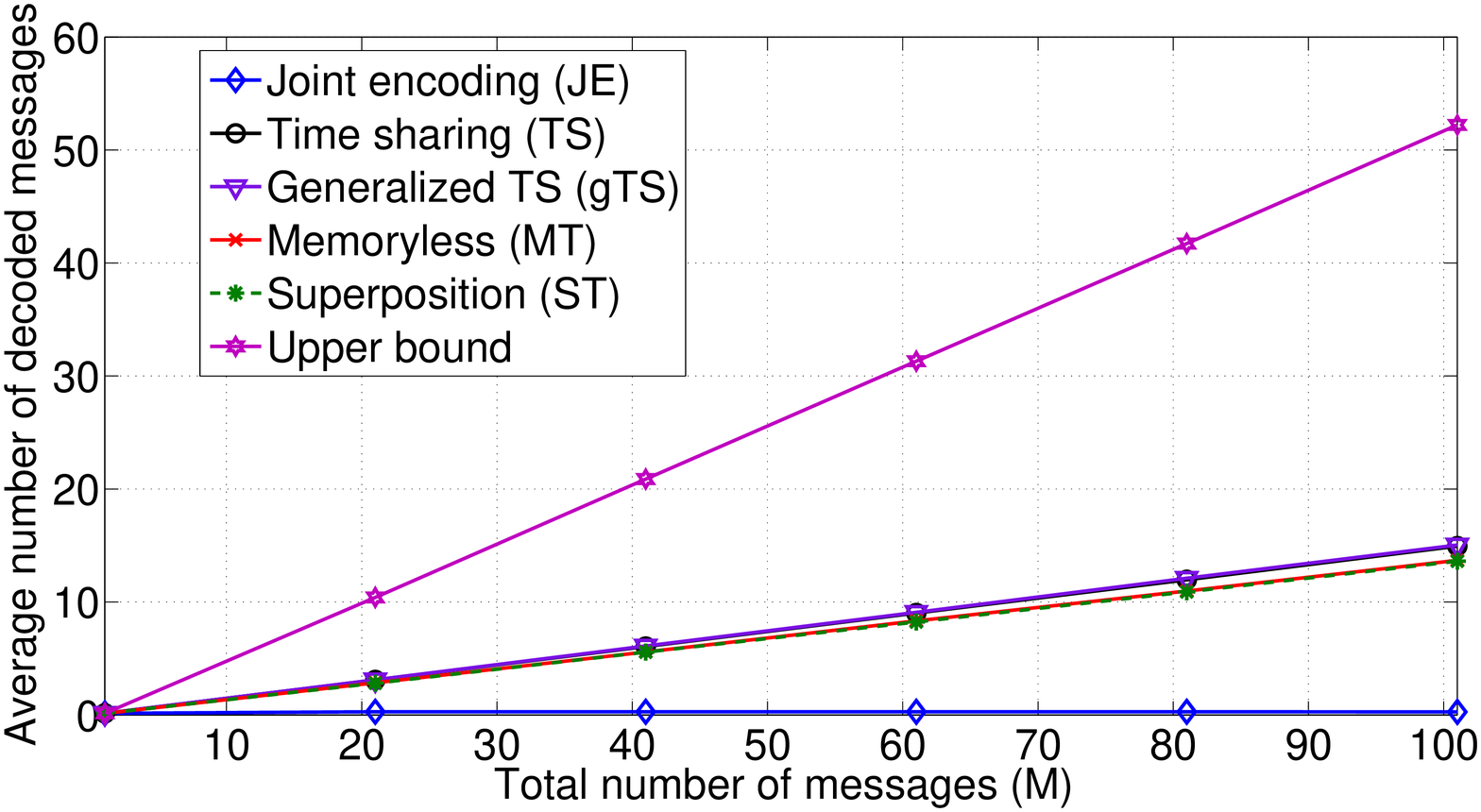}
        }%
        \subfigure[$P=2$
$\mathrm{dB}$ ($R<\overline{C}$).]{%
           \label{fig:rate_vs_M_2_db}
           \includegraphics[width=0.5\textwidth]{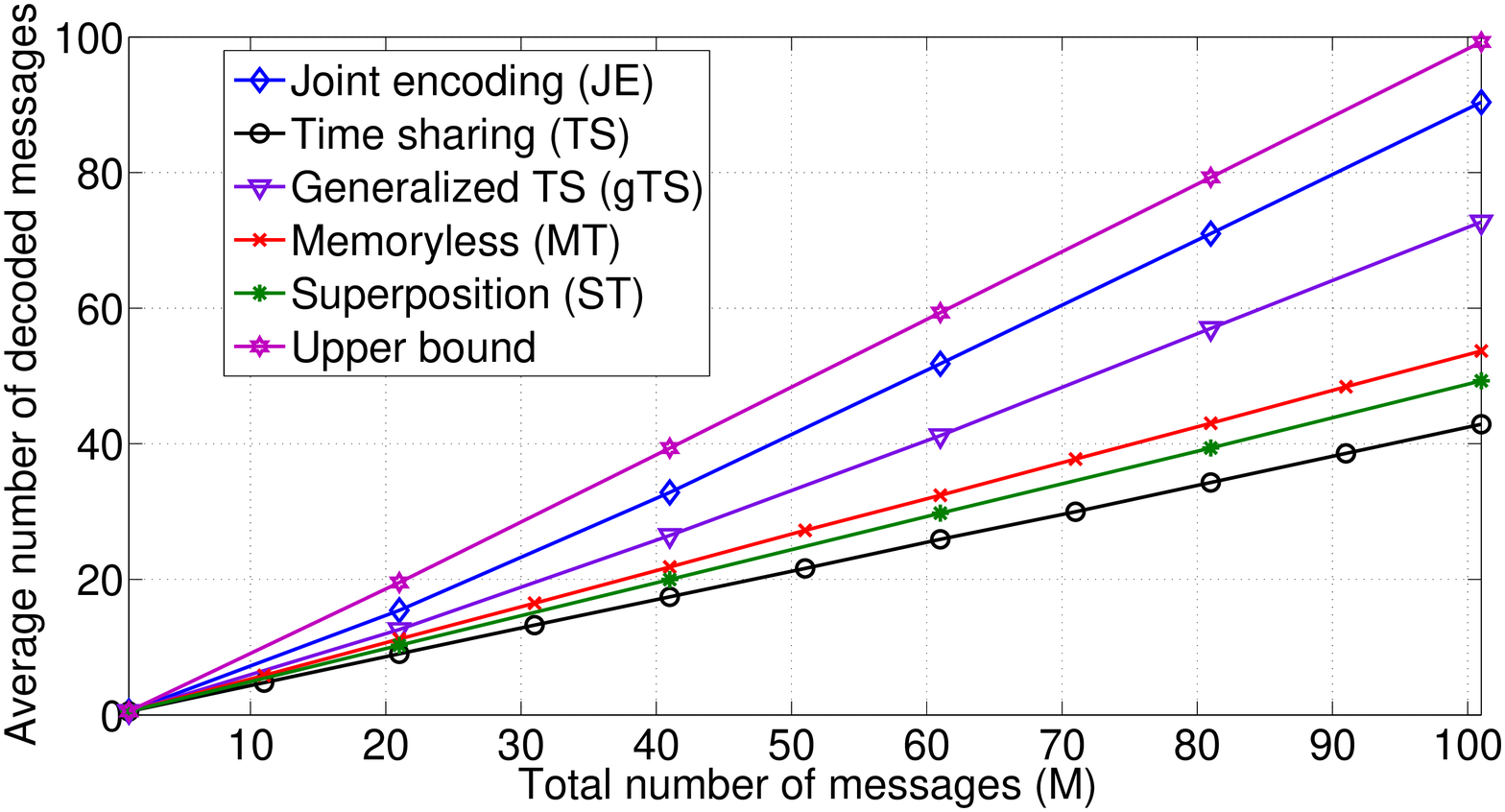}
        }\\
    \end{center}
    \caption{%
        Average number of decoded messages vs. the total number of messages $M$ for $R=1$ bpcu.
     }%
   \label{fig:subfigures}
\end{figure}
In Fig. \ref{fig:rate_vs_M_0_db} and Fig. \ref{fig:rate_vs_M_2_db}
the average number of decoded messages is plotted against $M$ for SNR values of $-3$ $\mathrm{dB}$ and
$2$ $\mathrm{dB}$, respectively, and a message rate of $R=1$ bpcu. While
JE outperforms the other schemes at $SNR=2$ $\mathrm{dB}$, it has the
poorest performance at $SNR=-3$ $\mathrm{dB}$. This behavior is
expected based on the threshold behavior of the JE scheme that we
have outlined in Section \ref{sec:joint}. Note that the average
capacity corresponding to $SNR=-3$ dB and $2$ dB are
$\overline{C}=0.522$ and $\overline{C}=1.158$, respectively. The
former is below the target rate $R=1$ and the receiver can not
decode almost any message, whereas the latter is above
$R=1$, leading to an average decoded rate close to the optimal value. Note
from the two figures that none of the schemes dominates the others at
all SNR values.
\begin{figure}[t!]
\centering
\includegraphics[width=4in]{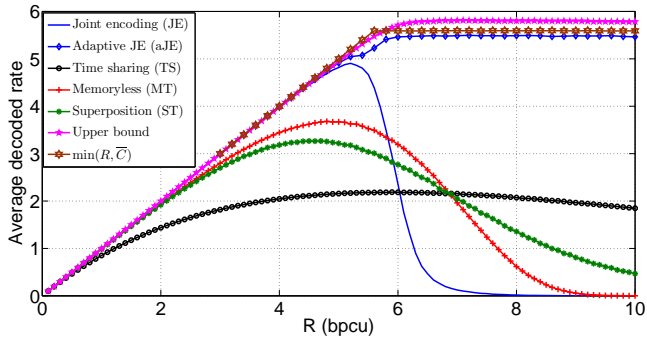}
\caption{Average decoded rate vs $R$ for $P=20$ $\mathrm{dB}$ and $M=100$ messages. The upper bound $\min(R,\overline{C})$ is also shown.}
\label{fig:variable_rate}
\end{figure}
In Fig. \ref{fig:variable_rate} $\overline{R}$ is plotted against the transmission rate $R$ for the case
of $M=100$ and $P=20$ $\mathrm{dB}$. The aJE scheme outperforms all the other schemes, performing very close to the upper bound. The number $M'$ of messages transmitted in the aJE scheme is chosen so that $\frac{M'}{M}=0.95\frac{\overline{C}}{R}$. In the figure we also show the upper bound obtained from the ergodic capacity $\min(R,\overline{C})$. It can be seen how it closely approximates the informed transmitter upper bound for $R<6$. The JE scheme performs better
than the others up to a certain transmission rate, beyond which
rapidly becomes the worst one. This is due to the phase transition behavior observed here even for a relatively small $M$. Among the other schemes, MT achieves the
highest average decoded rate in the region $R<6.8$,
while TS has the worst performance. The opposite is true in the
region $R>6.8$, where the curve of ST scheme is upper and lower
bounded by the curves of the MT and TS schemes. We have repeated the
simulations with different parameters (i.e., changing $P$ and $M$)
with similar results, that is, MT, TS, and ST schemes meet
approximately at the same point, below which MT has the best
performance of the three while above the intersection TS has the
best performance. At the moment we have no analytical explanation
for this observation, which would mean that there is always a scheme outperforming ST.
We next study the performance of the considered schemes as a function of the distance from the transmitter.
We scale the average received power
at the receiver with $d^{-\alpha}$, where $d$ is the distance from
the transmitter to the receiver and $\alpha$ is the path loss exponent.
The results are shown in Fig. \ref{fig:distance_comparison} for $P=20$ dB, $M=100$, $R=1$ bpcu and a path loss exponent $\alpha=3$.
The dependence of $\overline{R}$ on the distance is important, for instance, in the context of broadcast transmission in cellular networks, in which case the receiving terminals may have different distances from the transmitter. In such a scenario the range of the
average channel SNR values at the receivers becomes important, and the transmitter should use a
transmission scheme that performs well over this range. For
instance, in a system in which all users have the same average SNR,
which is the case for a narrow-beam satellite system where the SNR
within the beam footprint has variations of at most a few
$\mathrm{dB}$'s on average \cite{maral_bousq_satcom_sys}, the transmission scheme should perform
well around the average SNR of the beam. A similar situation may
occur in a microcell, where the relatively small radius of the cell
implies a limited variation in the average SNR range experienced by
the users at different distances from the transmitter. Instead, in the case
of a macrocell, in which the received SNR may vary significantly
from the proximity of the transmitter to the edge of the cell, the transmitter should
adopt a scheme which performs well over a larger range of SNR
values.
In the range up to $d=4$ the JE
scheme achieves the highest average decoded rate while
for $d\geq6$ the TS scheme outperforms the others. The drop in the decoded
rate in the JE scheme when passing from $d=4$ to $d=5$ is
similar to what we observe in Fig. \ref{fig:variable_rate} when the
rate increases beyond $R=6$ bpcu. In both cases the transition takes
place as the transmission rate surpasses the average channel
capacity.
The aJE scheme, which selects the fraction of messages to transmit based on $\overline{C}$, outperforms all other schemes and gets relatively close to the informed transmitter upper bound and the ergodic capacity. The aJE scheme adapts the average transmission rate at message level to the average channel capacity. We recall that, in the aJE scheme, the transmitter only has a statistical knowledge of the channel, and yet gets pretty close to the performance of a genie-aided transmitter even for a reasonably low number of channel blocks. We further notice how the adaptive JE scheme closely approaches the ergodic capacity, even though data arrives gradually at the transmitter during the transmission, instead of being available at the beginning, which is generally assumed for the achievability of the ergodic capacity \cite{Goldsmith97_capac_fading}.
We should note that in Fig. \ref{fig:variable_rate} the average transmission rate is optimized for each given distance for the aJE scheme, while such optimization is not done for the other schemes. Thus, in case two (or more) terminals have different distances from the transmitter, the optimization can no longer be performed and a tradeoff between the average decoded rates of the two nodes would be needed. The performance can be improved by considering a combination of the aJE scheme with the TS or ST schemes. The plots in Fig. \ref{fig:variable_rate} show how TS, MT and ST schemes are more robust compared to the JE scheme, as their average decoded rate decreases smoothly with the distance, unlike the JE scheme, which has a sudden drop.
\section{Conclusions}\label{sec:conclusions}
We have considered a transmitter streaming data to a receiver over a block fading channel,
such that the transmitter is provided with an independent message at
a fixed rate at the beginning of each channel block. We have used
the average decoded rate as our performance metric. We have proposed several new transmission schemes based on joint encoding, time-division and superposition encoding. A general upper bound on the
average decoded rate has also been introduced assuming the
availability of CSI at the transmitter.

We have shown analytically that the joint encoding (JE) scheme has a
threshold behavior and performs well when the target
rate is below the average channel capacity $\overline{C}$, while its performance drops sharply
when the target rate surpasses $\overline{C}$. To adapt to an average channel capacity that is below the fixed message rate $R$, the adaptive joint encoding (aJE) scheme transmits only some of the messages. We have proved analytically that the aJE scheme is asymptotically optimal as the number of channel blocks goes to infinity, even though data arrives gradually over time at a fixed rate, rather than being available initially. We have also shown numerically that, even for a finite number of messages, the aJE scheme outperforms other schemes in all the considered settings and performs close to the upper bound.

 We have also proposed the time-sharing (TS) and superposition transmission (ST) schemes, as well as a generalized TS scheme which transmits each message over a certain number of channel blocks. While none of these schemes outperform others at all settings, their performances degrade gradually with
the decreasing average SNR as opposed to the threshold behavior of the JE scheme. This provides robustness in the case of multiple receivers with different average SNRs or when the channel statistics information at the transmitter is not accurate.
\section*{Appendix}\label{sec:appendix}
\subsection{Proof of Theorem 1}
Let $B_k$ denote the event ``the first $k$ messages can be decoded
at the end of channel block $k$", while $\overline{B}_k$ denotes the
complementary event. The event $B_k$ holds if and only if
\begin{align}
    C_{k-i+1} + C_{k-i+2} + \cdots + C_k &\geq iR
\end{align}
is satisfied for all $i=1, \ldots,k$. Let $E_{k,j}$ denote the event ``the $j$-th inequality needed to decode the first $k$ messages in $k$ channel blocks is satisfied",
that is:
\begin{eqnarray}\label{eqn:proof_2}
E_{k,j}\triangleq\{C_{k-j+1}+\cdots+C_k\geq jR\},\mbox{\ for\ }
j=1,\ldots,k,
\end{eqnarray}
while $\overline{E}_{k,j}$ denotes the complementary event.

Note that in the JE scheme if $m$ messages are decoded these
are the first $m$ messages. Let $n_d$ denote the number
of decoded messages at the end of channel block $M$. Then the
average decoded rate is
\begin{eqnarray}\label{eqn:proof_4}
\overline{R}_{JE}=R\left[Pr\{n_d\geq1\}+Pr\{n_d\geq2\}+\cdots+\Pr\{n_d\geq
M-1\}+\Pr\{n_d\geq
M\}\right].
\end{eqnarray}
The $k$-th term in the sum of Eqn. (\ref{eqn:proof_4}) is the probability of
decoding \emph{at least} $k$ (i.e. $k$ or more) messages. Each term
in (\ref{eqn:proof_4}) can be expressed as the sum of two terms as:
\begin{eqnarray}\label{eqn:proof_5}
 Pr\{n_d\geq k\}=Pr\{B_k,n_d\geq k\}+Pr\{\overline{B}_k,n_d\geq k\}
\end{eqnarray}
The first term of the sum in (\ref{eqn:proof_5}) is the probability
of ``decoding $k$ messages at the end of
channel block $k$ \emph{and} decoding at least $k$ messages at the end of $M$
channel blocks". Note that this corresponds to event
$B_k$, since if $B_k$ holds, the event ``decode at least $k$
messages at the end of channel block $M$" is satisfied. We have:
\begin{eqnarray}\label{eqn:proof_6}
Pr\{B_k,n_d\geq k\}=Pr\{B_k\}=Pr\{E_{k,1},\cdots,E_{k,k}\}.
\end{eqnarray}
As for the second term of the sum in (\ref{eqn:proof_5}), it is the
probability of decoding at least $k$ messages but \emph{not} $k$ at
the end of channel block $k$. It can be further decomposed into the
sum of two terms, one corresponding to the probability of decoding and the other to the probability
of not decoding $k+1$ messages at the end of block $k+1$ while decoding
more than $k$ messages in $M$ blocks, i.e.:
\begin{eqnarray}\label{eqn:proof_7}
 Pr\{\overline{B}_k,n_d\geq k\}=Pr\{\overline{B}_k,B_{k+1},n_d\geq
 k\}+Pr\{\overline{B}_k,\overline{B}_{k+1},n_d\geq k\}.
\end{eqnarray}
Looking at the first term, similarly as seen before, the event
$n_d\geq k$ is true if the condition $B_{k+1}$ is satisfied (i.e.,
if $k+1$ messages are decoded at the end of block $k+1$, then more
than $k$ messages are decoded at the end of channel block $M$), that
is:
\begin{eqnarray}\label{eqn:proof_8}
 Pr\{\overline{B}_k,B_{k+1},n_d\geq
 k\}=Pr\{\overline{B}_k,B_{k+1}\}. \notag
\end{eqnarray}
Plugging these into (\ref{eqn:proof_5}), we obtain
\begin{eqnarray}\label{eqn:proof_x}
 Pr\{n_d\geq k\}=Pr\{B_k\}+Pr\{\overline{B}_k,B_{k+1}\}+Pr\{\overline{B}_k,\overline{B}_{k+1},n_d\geq
 k\}.
\end{eqnarray}
We can continue in a similar fashion, so that, in general the event ``at
least $k$ messages are decoded" can be written as the union of the
disjoint events (``$k$ messages are decoded in $k$ slots") $\bigcup$
(``k messages are not decoded in $k$ slots but $k+1$ messages are
decoded in $k+1$ slots") $\bigcup$ $\cdots$ $\bigcup$ (``no message
can be decoded before slot $M$ but $M$ messages are decoded in slot
$M$"). Hence, by the law of total probability, the probability of
decoding more than $k$ messages can be written as:
\begin{eqnarray}\label{eqn:proof_9}
 Pr\{n_d\geq k\}=\sum_{j=k}^MPr\{\overline{B}_k,\overline{B}_{k+1},\cdots,\overline{B}_{j-1},B_{j}\}.
\end{eqnarray}
Note that each term of the sum in (\ref{eqn:proof_9}) says nothing
about what happens to messages beyond the $j$-th, which can either
be decoded or not. Plugging (\ref{eqn:proof_9}) in
(\ref{eqn:proof_4}) we find:
\begin{eqnarray}\label{eqn:proof_9bis}
E[m]=\sum_{k=1}^MPr\{n_d\geq
k\}=\sum_{k=1}^M\sum_{j=k}^MPr\{\overline{B}_k,\overline{B}_{k+1},\cdots,\overline{B}_{j-1},B_j\}\notag\\
=\sum_{j=1}^M\sum_{k=1}^jPr\{\overline{B}_k,\overline{B}_{k+1},\cdots,\overline{B}_{j-1},B_j\}.
\end{eqnarray}

We can rewrite each of these events as the intersection of
events of the kind $E_{k,i}$ and $\overline{E}_{k,i}$. Each term of
the sum in (\ref{eqn:proof_9bis}) can be split in the
sum of the probabilities of two disjoint events:
\begin{eqnarray}\label{eqn:proof_10}
Pr\{\overline{B}_k,\overline{B}_{k+1},\cdots,\overline{B}_{j-1},B_j\}=
Pr\{E_{k,1},\overline{B}_k,B_{k+1},\cdots,\overline{B}_{j-1},B_j\}\notag\\+
Pr\{\overline{E}_{k,1},\overline{B}_k,\overline{B}_{k+1},\cdots,\overline{B}_{j-1},B_j\}.
\end{eqnarray}
As the event $\overline{E}_{k,1}$ implies the event
$\overline{B}_k$, this can be removed from the second term in the
right hand side of (\ref{eqn:proof_10}). Note that, in general, the
event $\overline{E}_{k,i}$, $i\in \{1,\cdots,k\}$ implies the event
$\overline{B}_k$. In order to remove the event $\overline{B}_k$ from
the first term as well, we write it as the sum of probabilities of
two disjoint events: one intersecting with $E_{k,2}$ and the other
with $\overline{E}_{k,2}$. Then we get:\pagebreak[4]
\begin{eqnarray}\label{eqn:proof_11}
Pr\{\overline{B}_k,\overline{B}_{k+1},\cdots,\overline{B}_{j-1},B_j\}=
Pr\{E_{k,1},E_{k,2},\overline{B}_k,\cdots,\overline{B}_{j-1},B_j\}\notag\\+
Pr\{E_{k,1},\overline{E}_{k,2},\overline{B}_k,\cdots,\overline{B}_{j-1},B_j\}\\+\notag
Pr\{\overline{E}_{k,1},\overline{B}_{k+1},\cdots,\overline{B}_{j-1},B_j\}.
\end{eqnarray}
Now $\overline{B}_k$ can be removed from the second term of the sum
thanks to the presence of $\overline{E}_{k,2}$.
 Each of the terms in the right hand side of (\ref{eqn:proof_11}) can be further
written as the sum of the probabilities of two disjoint events and
so on so forth. The process is iterated until all the
$\overline{B}_d$, $d<j$ events are eliminated and we are left with events that are intersections of only events of
the type $E_{p,q}$ and $\overline{E}_{p,q}$, for some
$p,q\in\{k, k+1,\ldots,M\}$ and $B_j$. The iteration is done as follows:

For each term of the summation, we take the $\overline{B}_l$ event
with the lowest index. If any $\overline{E}_{l,j}$ event is present,
then $\overline{B}_l$ can be eliminated. If not, we write the term as
the sum of the two probabilities corresponding to the events which
are the intersections of the $\overline{B}_l$ event with $E_{l,d+1}$
and $\overline{E}_{l,d+1}$, respectively, where $d$ is the highest
index $j$ among the events in which $E_{l,j}$ is already present.
The
iterative process stops when $l=j$.

At the end of the process all the probabilities involving events
$\overline{B}_k, \ldots ,\overline{B}_{j-1}$ will be removed and
replaced by sequences of the kind:
$$\{E_{k,1},E_{k,2},\cdots,\overline{E}_{k,i_k},
E_{k+1,i_k+1},\cdots,\overline{E}_{k+1,i_{k+1}},\cdots,E_{j-1,i_{j-2}+1},\overline{E}_{j-1,i_{j-1}},B_j\},$$
where $i_{j-1}\in\{j-1-k,\cdots,j-1\}$ is the index corresponding to
the last inequality needed to decode $j-1$ messages which is not
satisfied. Note that exactly one $\overline{E}_{l,r}$ event for each
$\overline{B}_l$ is present after the iteration.

In order to guarantee that $B_j$ holds, all the events
$E_{j,1},\ldots,E_{j,j}$ must be verified. It is easy to show that,
after the iterative process used to remove the $\overline{B}_l$'s, the event $E_{j,i_{j-1}+1}$ ensures that all the events
needed for $B_j$ with indices lower than or equal to $i_{j-1}$ are
automatically verified. Thus, we can add the events
$\{E_{j,i_{j-1}+1},\cdots,E_{j,j}\}$ to guarantee that $B_j$ holds,
and remove it from the list. It is important to notice that the term
$E_{j,j}$ is always present. At this point we are left with the sum
of probabilities of events, which we call \emph{$E$-events}, each of
which is the intersection of events of the form $E_{i,j}$ and
$\overline{E}_{i,j}$. Thus, an $E$-event $S_k^j$ has the following form:
\begin{eqnarray}\label{eqn:eta_event}
S^j_k\triangleq\{E_{k,1},E_{k,2},\cdots,\overline{E}_{k,i_k},
E_{k+1,i_k+1},\cdots,\overline{E}_{k+1,i_{k+1}},\cdots,E_{j-1,i_{j-2}+1},\overline{E}_{j-1,i_{j-1}},E_{j,i_{j-1}+1},\cdots,E_{j,j}\}.
\end{eqnarray}
By construction, the number of $E$-events
for the generic term $j$ of the sum in (\ref{eqn:proof_9bis}) is
equal to the number of possible dispositions of $j-k$
$\overline{E}$'s over $j-1$ positions. As the number of events of
type $\overline{E}$ is different for the $E$-events of different
terms in (\ref{eqn:proof_9bis}), the $E$-events relative to two
different terms of (\ref{eqn:proof_9bis}) are different. We define
$\mathcal{S}_j$ as the set of all $E$-events which contain the event
$E_{j,j}$. The elements of $\mathcal{S}_j$ correspond to all the
possible ways in which $j$ messages can be decoded at the end of
block number $j$. The cardinality of $\mathcal{S}_j$ is equal to:
\begin{eqnarray}\label{eqn:proof_12}
\left|\mathcal{S}_j\right|=\sum_{k=1}^j\frac{(j-1)!}{(k-1)!(j-k)!}=2^{j-1},
\end{eqnarray}
which is the number of all possible combinations of $j-1$
elements each of which can take value $E$ or $\overline{E}$. Now we
want to prove that
\begin{align}\label{eqn_set_sum_proof}
\sum_{S^j_k\in\mathcal{S}_j}Pr\{S^j_k\}=Pr\{E_{j,j}\}.\end{align}
 Note that
$E_{k,l}$'s correspond to different events if the index $k$ is
different, even for the same index $l$; thus, the law of total
probability can not be directly applied to prove (\ref{eqn_set_sum_proof}).
However, the following can be easily verified:
$Pr\{E_{k_1,l}\}=Pr\{E_{k_2,l}\}$, $\forall k_1,k_2$. This implies
that the probabilities of two $E$-events which differ in some or all
of the $k$ indices (but not in the $l$ indices) of its constituent
events are the same. A proof is given in the following.

%

\emph{Proposition 1:} Let us consider a set of random variables $C_1,\cdots,C_j$ that are conditionally i.i.d. given $U$. Given any two ordering vectors $\mathbf{i}=i_1,i_2,\cdots,i_j$ and $\mathbf{l}=l_1,l_2,\cdots,l_j$, we have
\begin{eqnarray}\label{eqn:propos_to_prove}
Pr\{C_{i_1}\gtrless R,\ldots,C_{i_1}+\cdots+C_{i_j}\gtrless jR\}=Pr\{C_{l_1}\gtrless R,\cdots,C_{l_1}+\cdots+C_{l_j}\gtrless jR\},\end{eqnarray}

\emph{Proof:} The left hand side of Eqn. (\ref{eqn:propos_to_prove}) can be rewritten as:
\begin{eqnarray}\label{eqn:propos_proof}
Pr\{C_{i_1}\gtrless R,\ldots,C_{i_1}+\cdots+C_{i_j}\gtrless jR\}=\int_{-\infty}^{+\infty}du \int_{\theta_1^{low}}^{\theta_1^{up}}dc_{i_1}
\dots \int_{\theta_j^{low}}^{\theta_j^{up}}dc_{i_j}f_{\mathbf{C}_{\mathbf{i}}|U}(\mathbf{c}_{\mathbf{i}}|u)f_{U}(u),\end{eqnarray}
where $\mathbf{C}_{\mathbf{i}}=C_{i_1},\cdots,C_{i_j}$ and $\mathbf{c}_{\mathbf{i}}=c_{i_1},\cdots,c_{i_j}$, while $\theta_h^{low}$ and $\theta_h^{up}$ are the lower and upper extremes of the integration interval. $\theta_h^{low}$ is either equal to $-\infty$ or to $hR-c_{i_1}-\cdots-c_{i_{h-1}}$, $\forall h \in \{1,\ldots, j\}$, depending on whether there is a $<$ or a $\geq$ in the $h$-th inequality within brackets in Eqn. (\ref{eqn:propos_proof}), respectively, while $\theta_h^{up}$ is either equal to $hR-c_{i_1}-\cdots-c_{i_{h-1}}$ or to $+\infty$ depending on whether there is a $<$ or a $\geq$ in the $h$-th inequality of Eqn. (\ref{eqn:propos_proof}), respectively.
By plugging Eqn (\ref{eqn:propos_condit_iid}) into Eqn (\ref{eqn:propos_proof}) we can write:
\begin{eqnarray}\label{eqn:propos_proof2}
Pr\{C_{i_1}\gtrless R,\ldots,C_{i_1}+\cdots+C_{i_j}\gtrless jR\}=\int_{-\infty}^{+\infty}du f_{U}(u)\int_{\theta_1^{low}}^{\theta_1^{up}}dc_{i_1}
\dots \int_{\theta_j^{low}}^{\theta_j^{up}}dc_{i_j}f_{\mathbf{C}_{\mathbf{i}}|U}(\mathbf{c}_{\mathbf{i}}|u)\notag\\
=\int_{-\infty}^{+\infty}du f_{U}(u)\int_{\theta_1^{low}}^{\theta_1^{up}}dc_{i_1}
\dots \int_{\theta_j^{low}}^{\theta_j^{up}}dc_{i_j}f_{C_{i_1}|U}(c_{i_1}|u)\cdots f_{C_{i_j}|U}(c_{i_j}|u).\end{eqnarray}
Finally, by using Eqn. (\ref{eqn:propos_condit_iid2}) in Eqn. (\ref{eqn:propos_proof2}) we find:
\begin{eqnarray}\label{eqn:propos_proof3}
&Pr\{C_{i_1}\gtrless R,\ldots,C_{i_1}+\cdots+C_{i_j}\gtrless jR\}=\int_{-\infty}^{+\infty}du f_{U}(u)\int_{\theta_1^{low}}^{\theta_1^{up}}dc_{i_1}
\dots \int_{\theta_j^{low}}^{\theta_j^{up}}dc_{i_j}f_{\mathbf{C}_{\mathbf{i}}|U}(\mathbf{c}_{\mathbf{i}}|u)\notag&\\
&=\int_{-\infty}^{+\infty}du f_{U}(u)\int_{\theta_1^{low}}^{\theta_1^{up}}dc_{i_1}
\dots \int_{\theta_j^{low}}^{\theta_j^{up}}dc_{i_j}f_{C_{i_1}|U}(c_{i_1}|u)\times\cdots \times f_{C_{i_j}|U}(c_{i_j}|u)\notag&\\
&=\int_{-\infty}^{+\infty}du f_{U}(u)\int_{\theta_1^{low}}^{\theta_1^{up}}dc_{l_1}
\dots \int_{\theta_j^{low}}^{\theta_j^{up}}dc_{l_j}f_{C_{l_1}|U}(c_{l_1}|u)\times\cdots \times f_{C_{l_j}|U}(c_{l_j}|u)\notag&\\&
=Pr\{C_{l_1}\gtrless R,\ldots,C_{l_1}+\cdots+C_{l_j}\gtrless jR\}.\blacksquare&
\end{eqnarray}
The proposition above guarantees that, although these events do not
partition the whole probability space of $E_{j,j}$, their
probabilities add up to that of $E_{j,j}$, i.e.:
\begin{eqnarray}\label{eqn:proof_13}
\sum_{k=1}^{2^{j-1}}Pr\{S^j_k\}=Pr\{E_{j,j}\}=Pr\{C_1+\cdots+C_j\geq
jR\}.
\end{eqnarray}
Finally, plugging Eqn. (\ref{eqn:proof_13}) into Eqn.
(\ref{eqn:proof_9bis}) we can write:
\begin{eqnarray}\label{eqn:proof_14}
E[m]=\sum_{k=1}^MPr\{n_d\geq k\}&=&
\sum_{j=1}^M\sum_{k=1}^jPr\{\overline{B}_k,\overline{B}_{k+1},
\cdots,\overline{B}_{j-1},B_j\}\notag\\&=&\sum_{j=1}^M\sum_{S^j_k\in
\mathcal{S}_j}Pr\{S^j_k\}=\sum_{j=1}^MPr\{C_1+\cdots+C_j\geq jR\}.
\blacksquare
\end{eqnarray}
\nopagebreak[4]
\bibliographystyle{ieeetran}
\bibliography{articoli}
\end{document}